\documentclass[smallextended]{svjour3}       % onecolumn (second format)
\smartqed  % flush right qed marks, e.g. at end of proof
\usepackage{graphicx,amssymb}
\usepackage{amsmath}

\numberwithin{equation}{section}
\journalname{Journal of Mathematical Biology}
\usepackage{natbib}
\begin{document}

\title{Coarse-grained analysis of stochastically simulated cell populations with a positive feedback genetic network architecture.%\thanks{Grants or 
}

\titlerunning{Coarse-grained analysis of stochastically simulated cell populations%\thanks{Grants or 
}        % if too long for running head

\author{I. G. Aviziotis         \and
        M. E. Kavousanakis 
        \and
        I.A. Bitsanis
        \and
        A. G. Boudouvis %etc.
}


\institute{I.G. Aviziotis \at
              National Technical University of
Athens, School of Chemical Engineering, Athens, 15780, Greece 
%              Fax: +123-45-678910\\
              \email{javiziot@chemeng.ntua.gr}           %  \\
           \and
M.E. Kavousanakis
\at
                         National Technical University of
Athens, School of Chemical Engineering, Athens, 15780, Greece\\
              Tel.: +30-772-3290\\
\email{mihkavus@chemeng.ntua.gr}
\and
I.A. Bitsanis 
\at
Institute of Electronic Structure and Laser, Foundation for Research and Technology Hellas, GR-711 10 Heraklion, Greece
\email{bitsanis@iesl.forth.gr}
\and
A.G. Boudouvis
\at
           National Technical University of
Athens, School of Chemical Engineering, Athens, 15780, Greece
\email{boudouvi@chemeng.ntua.gr}
}


\maketitle

\begin{abstract}
Among the different computational approaches modelling the dynamics of isogenic cell populations, discrete stochastic models can describe with sufficient accuracy the evolution of small size populations.
However, for a systematic and efficient study of their long-time behaviour over a wide range of parameter values, the performance of solely direct temporal simulations requires significantly high computational time.
In addition, when the dynamics of the cell populations exhibit non-trivial bistable behaviour, such an analysis becomes a prohibitive task, since a large ensemble of initial states need to be tested for the quest of possibly co-existing steady state solutions.
In this work, we study cell populations which carry the {\it lac} operon network exhibiting solution multiplicity over a wide range of extracellular conditions (inducer concentration).
By adopting ideas from the so-called ``equation-free'' methodology, we perform systems-level analysis, which includes numerical tasks such as the computation of {\it coarse} steady state solutions, {\it coarse} bifurcation analysis, as well as {\it coarse} stability analysis.   
Dynamically stable and unstable macroscopic (population level) steady state solutions are computed by means of bifurcation analysis utilising short bursts of fine-scale simulations, and the range of bistability is determined for different sizes of cell populations. 
The results are compared with the deterministic cell population balance (CPB) model, which is valid for large populations, and we demonstrate the increased effect of stochasticity in small size populations with asymmetric partitioning mechanisms. 

\keywords{{\it lac} operon \and Monte Carlo simulations \and equation-free method \and bistability \and cell heterogeneity}
\subclass{92D25 \and 93E03 \and 9208}
\end{abstract}

\section{Introduction}
\label{intro}

Molecular biology, genomics, transcriptomics and proteomics have provided us the
appropriate powerful tools for the investigation and understanding of the immensely complex processes, which occur at the single-cell level.
However, the biological behavior is not a matter of solely intracellular processes; 
it also depends on the interactions, which take place between the cells of an isogenic population, leading to a phenotypic variability amongst them.
This phenomenon is known as cell population heterogeneity and has been observed in an abundance of biological systems,
e.g., phage burst size variations \cite{Delbruck:1945:BSD}, transcriptional states heterogeneity in sporulating cultures of {\it Bacillus subtilis} \cite{Chung:1995:STH}, heterogeneity in endothelial cell surface markers \cite{Oh:2004:SPM} and in various isogenic {\it Escherichia coli} systems \cite{Elowitz:2002:SGE}.
Until the previous decade, the biological paradigms and modeling frameworks for the design and control of biochemical processes did not consider cell population heterogeneity \cite{Chung:1995:STH,Fedorrof:2002}.
Their key assumption was that all cells of an isogenic population behave like the average cell, and simple ordinary differential equations were used to describe the population behavior \cite{Avery:2006,Davidson:2008}.
However, such an approach can result in incorrect predictions as shown in \cite{McAdams:1998,Mantzaris:2005:CPB,Kavousanakis:2009} and one must account for the heterogeneous nature of cell populations.
Furthermore, much of our knowledge is based on ensemble measurements, and despite the fact that cell-to-cell differences are always present to some degree, the collective behaviour of a population may not represent the behaviour of the individuals \cite{Altschuler:2010}.
The accurate prediction of the collective behaviour of cell populations is of great importance in the majority of biotechnological applications, where the objective is to maximise their productivity, rather than increasing the efficiency of each individual cell.
Furthermore, phenotypic heterogeneity may be linked to the viability of a cell population and its ability to adapt to abrupt changes of their environment \cite{McAdams:1999,Sumner:2002,Veening:2008}.
Thus, the resistance of certain infectious bacteria to antibiotics could be explained on the basis of the existence of small sub-populations that survive the medical treatment and resume growing after the antibiotic has been removed \cite{Booth:2002}.
From the discussion above, it naturally emerges that a mathematical description that incorporates heterogeneity in cell population dynamics is of great importance
for a variety of biological systems.
Fredrickson and coworkers introduced a special class of models in the 1960s for the prediction of the behavior of heterogeneous cell populations \cite{Eakman:1966:SDP,Tsuchiya:1966:DMP,Fredrickson:1967:SDP}, known as Cell Population Balance (CPB) models.
These models are partial integro-differential equations, which describe the physiological state of cells, i.e., a vector whose components can be the intracellular content of chemical species, as well as morphometric characteristics of the cells, e.g., size.
CPB models incorporate the type of heterogeneity originating from the unequal partitioning of most intracellular components (with the exception of DNA) between the daughter cells \cite{Block:1990}.
Due to the operation of the cell cycle, this phenomenon repeats itself, thus leading to further variability.
However, the validity of CPB models is limited only when the continuum assumption is justified, i.e., for large size populations. 
Furthermore, the deterministic approach does not account for stochastic effects during division, which can play a key role in small size cell populations.
For these reasons, and in order to provide an -as much as possible- accurate description of the system behaviour, we resort to fine-scale, stochastic modelling approaches.
The first modelling approach for capturing the stochastic behavior of cell populations can be found in \cite{Shah:1976}, who developed a Monte Carlo algorithm to describe the dynamics of the cell mass distribution.
In Hatzis et al. (1995), the Monte Carlo algorithm was extended to describe a system of increased complexity describing the growth of phagotrophic protozoa.
These algorithms are computationally intensive due to the exponential growth in time of the number of cells in the population.
Constant-number Monte Carlo (CNMC) algorithms \cite{Smith:1998,Mantzaris:2006:SDS} overcome the extensive requirements in CPU time, by simulating a constant number of cells that are assumed to be a representative sample of the overall population.
Despite the fact that stochastic modelling provides a more realistic description of the physical problem, their computational efficiency is limited when a coarse-level analysis is required. 
For example, if we are interested in studying the asymptotic behaviour of a cell population with respect to certain parameters (e.g., extracellular inducer concentration), then a large number of dynamic -long time- stochastic simulations is required. 
Furthermore, for cases where the system is expected to feature solution multiplicity within a range of parameter values, the proper initial states need to be chosen in order to converge to all possible steady state solutions, with no guarantee of success since the limits of bistability are not known {\it a priori}.
For a systematic and efficient analysis of the cell population coarse behaviour, we adopt ideas from the so-called ``equation-free'' methodology \cite{Gear:2002,Kevrekidis:2003,Kevrekidis:2004}, a computer-assisted multiscale framework that enables models at a fine (microscopic) level of description to perform numerical tasks at a coarse (macroscopic) level.
The backbone of the methodology is the development of a computational super-structure, which wraps around the microscopic simulator and reports the evolution of coarse quantities of interest at discrete time instances. 
This discrete time-mapping, which is called the {\it coarse} time-stepper, allows the communication between the single-cell and population level.
Using as a tool the coarse time-stepper, we implement well established numerical techniques for the performance of systems-level analysis tasks, such as the computation of coarse steady-state solutions, coarse bifurcation analysis and coarse stability analysis.
Here, we demonstrate the efficiency of this multiscale methodology, when the fine-scale, microscopic simulator is a CNMC model  \cite{Mantzaris:2006:SDS}, which simulates the evolution of an isogenic cell population carrying the {\it lac} operon genetic network.
Within a range of extracellular inducer concentration values (IPTG), the {\it lac} operon exhibits bistability, i.e.,
for the same parameter value asymptotic cell phenotypes of high or low expression levels of the {\it lac}Y gene can be both observed.
At the single-cell level, the solution multiplicity is attributed to the nonlinear expression of the rate of reaction in which the intracellular content ({\it lac}Y) participates \cite{Mantzaris:2007:FSP}. 
The same behaviour is also inherited to the population level, when solving a deterministic CPB model, which incorporates cell heterogeneity\cite{Kavousanakis:2009}. 
In fact, cell heterogeneity has a significant impact on the average population phenotype shifting the bistability region towards higher IPTG concentration values.
The accurate determination of the bistability region limits is realised by applying bifurcation analysis, and in particular by means of the pseudo arc-length parameter continuation technique \cite{Keller:1977}.
In this work, we perform the same parametric analysis using a stochastic model, in order to study the effect of random events during division on the limits of the bistability range.  
The performance of such systems-level analysis is feasible by means of the equation-free methodology.
In particular, we perform coarse steady state computations and coarse bifurcation analysis, utilising CNMC simulations \cite{Mantzaris:2006:SDS}.
This study enables the determination of the bistability range for different population sizes, which is compared with results obtained from the numerical solution of deterministic CPB model.
Major differences are observed for small size cell populations, with highly asymmetric partitioning mechanism;
the computed bistability region for small size populations is located within the range of higher IPTG concentration values compared to the deterministic analysis findings. 
The accurate determination of the bistability region becomes a valuable tool for the understanding of possible phenotypic switching induced by random events \cite{Libby:2011}.
When the cell population operates at the proximity of the bistability interval limits, small fluctuations can drive the system to large phenotypic changes.
Here, we illustrate switching from high to low (and reverse) levels of {\it lac}Y gene expressions, when the parameter value (here the IPTG concentration) is in the vicinity of critical turning points, which are computed from the coarse bifurcation analysis. 
This paper is organised as follows: In Section~\ref{sec:kmcmodel}, we present the CNMC model, which simulates the evolution of an isogenic cell population.
We briefly describe the {\it lac} operon network and present results of direct stochastic simulations for different IPTG concentrations. 
The equation-free methodology is described in Section~\ref{sec:coarsemethod};
in Section~\ref{sec:results}, we present results of the proposed coarse-grained analysis, studying the effect of partitioning asymmetry, and of the division rate sharpness for different sizes of cell populations.
Finally, we summarise the main findings of this study in Section~\ref{sec:conclusions} and propose future research directions.

\section{Model description}
\label{sec:kmcmodel}

\subsection{Stochastic model}
The stochastic model is a CNMC algorithm, which accounts for stochastic events during cell division \cite{Mantzaris:2006:SDS}.
All individuals carry the same gene regulatory network (e.g., {\it lac} operon), and we denote with $S_\tau$ the random state of the cell population at a given dimensionless time $\tau$:
\begin{equation}
\label{eq:randomstate}
S_{\tau} \equiv \left \{  X_i(\tau) = x_i, i=1,2,...,N \right \},
\end{equation}
\noindent where $X_i(\tau)$ is the intracellular content of cell $i$, and $N$ is the constant number of cells of the simulated population.
Each cell undergoes division at a rate $\Gamma(x)$ and a partition probability density function $P(x,x')$ determines the intracellular content of newborn cells.
Between division events, the content of each cell changes according to an ordinary differential equation, the exact formulation of which depends on the genetic network the cells carry:

\begin{equation}
\label{eq:react}
\frac{\textrm{d}x_i}{\textrm{d}\tau}=R(x_i), \quad i=1,2,...,N.
\end{equation}

\subsection*{\it Computation of time between division events}
The time between division events, $T$, is a random variable which depends on $S_\tau$;
its cumulative distribution function is given from \cite{Mantzaris:2006:SDS}:

\begin{equation}
\label{eq:div_time}
F_{\tau}(z|\tau) = 1-\exp \left[ -\int_0^z {\sum_{i=1}^{N} \Gamma \left( x_i(\tau+z')  \right)} \textrm{d} z' \right].
\end{equation}

\noindent In order to determine the time between two successive division events, a random number $p_1$ is then generated from a uniform distribution, with $p_1 \in [0,1]$ and we solve the nonlinear equation:

\begin{equation}
\label{eq:findtime}
\frac{ \sum_{i=1}^{N} \int_{0}^{T} \Gamma \left( x_i(\tau+z') \right) \textrm{d}z'}{\textrm{ln} \left[1-p_1 \right]} +1 =0.
\end{equation}
Equation (\ref{eq:findtime}) is solved iteratively with the Newton-Raphson method;
the trapezoid rule is applied for the numerical computation of the integral.
The evolution of cell's $i$ intracellular content, $x_i$, during this intervening time period is computed by integrating in time Eq. (\ref{eq:react}) with the explicit forward Euler scheme.
Alternative time integration schemes can be also applied, however with no significant effect on the results of simulations.

\subsection*{\it Determination of the cell undergoing division}

After the computation of time $T$ between division evens, the state vector $S_\tau$ is updated by integrating Eq. (\ref{eq:react}) (with the forward Euler scheme)
and the cell undergoing division is selected.
The dividing cell (denoted with index $k$) is randomly selected from the conditional distribution function:
\begin{equation}
\label{eq:divcell}
\textrm{Pr} \left \{ k=j | S_{\tau+T} \right \} = \frac{ \Gamma \left( x_j (\tau+T) \right)}{\sum_{i=1}^{N} \Gamma \left( x_i (\tau+T) \right)}.
\end{equation}

\subsection*{\it Determination of the content of the two daughter cells}

Upon division of cell $k$, the content of mother cell  is partitioned to the newborn cells according to a partition probability density function $P(x,x')$,
which determines the mechanism through which a mother cell of content $x'$ produces a daughter cell with content
$x$ and a second cell with content $x'-x$.
In the results presented below, we choose to work with the simple discrete partitioning mechanism, formulated as:

\begin{equation}
\label{eq:partition}
P(x,x') = \frac{1}{2f} \delta(fx'-x)+ \frac{1}{2(1-f)} \delta \left( (1-f)x'-x \right),
\end{equation}
\noindent where $\delta$ is the Dirac function and $f$ is an asymmetry parameter.
In practice, the division of the mother cell with intracellular content $x'$ results in two daughter cells with content $fx'$ and $(1-f)x'$, respectively.
From this definition, it is clear that $f \in [0,0.5]$, with low values of $f$ corresponding to more asymmetric intracellular content partitioning.
When $f=0.5$ the partitioning mechanism is symmetric.

\subsection*{\it Restoration of the sample}

Each division event leads to the generation of two newborn cells from the mother cell.
Upon determination of the content of each of the daughter cells, the mother cell is replaced by the first daughter cell.
In order to maintain the size of population constant at a pre-specified value, $N$, the second daughter cell replaces
a randomly chosen cell from the population, with all cells having equal probability of being selected and replaced.
Finally, the dimensionless time is updated, $\tau \rightarrow \tau + T$, and the algorithm is repeated until a pre-specified stoppage time is reached.

\subsection{Deterministic model}
The corresponding deterministic description is provided by the CPB model \cite{Mantzaris:2005:CPB,Kavousanakis:2009}:

\begin{eqnarray}
\label{eq:determ}
\nonumber \frac{\partial n(x,t)}{\partial t} + \frac{\partial}{\partial x} \left[ R(x) n(x,t) \right] + \Gamma(x) n(x,t)= \\
= 2 \int_x^{x_{max}} \Gamma(x')P(x,x')n(x',t)\textrm{d}x' - n(x,t) \int_0^{x_{max}} \Gamma(x)n(x,t) \textrm(d)x,
\end{eqnarray}

\noindent where $n(x,t)$ is the cell density function and denotes the number of cells with content $x$ at time $t$, divided by the total number of cells at the same time.
The maximum intracellular content value is denoted with $x_{max}$.
The boundary conditions imposed to (\ref{eq:determ}) require that the cells of the population do not grow outside the domain $[0,x_{max}]$: 
\begin{equation}
n(0,t)=n(x_{max},t) = 0.
\end{equation}

If we take the first-order moment of (\ref{eq:determ}) and apply the mass conservation of the intracellular component at cell division then the average intracellular content $\left< x \right>$ is computed from:
\begin{equation}
\label{eq:average}
\frac{\textrm{d} \left< x \right>}{\textrm{d} t} = \int_0^{x_{max}} R(x) n(x,t) \textrm{d}x  - \left< x \right> \int_0^{x_{max}} \Gamma(x) n(x,t) \textrm{d} x.
\end{equation}
For the numerical solution of the deterministic cell population balance model (partial integro-differential equation) a moving boundary algorithm was developed and applied in Kavousanakis et al. (2009).
The dynamics of (\ref{eq:determ}) -as well as of the stochastic model- are fully determined through the definition the division rate $\Gamma(x)$, the partitioning mechanism $P(x,x')$ and the single-cell reaction rate $R(x)$, which contains the intracellular species described in the model. 
These three functions are generally known as intrinsic physiological state functions (IPSF).
For the division rate, we consider the following normalised power law \cite{Dien:1994:CDS}:
\begin{equation}
\label{eq:division}
\Gamma(x) = \left( \frac{x}{\left<x\right>} \right)^m,
\end{equation}
\noindent where $m$ controls the sharpness of $\Gamma$.
Large $m$ values correspond to  sharper division rates, i.e. a larger separation of rates under which cells below and above a certain intracellular content can undergo division.
The partitioning of intracellular content upon division is discrete following the expression (\ref{eq:partition}).

The expression of the single-cell reaction rate $R(x)$ depends on the genetic network carried by the cells of the population. 
In this work, we study  a class of regulatory genetic networks with positive feedback architecture, and in particular the {\it lac} operon, which constitutes one of the most well-studied gene regulatory networks \cite{Beckwith:1970:TLO,Miller:1978:TO}.

\section{A genetic network with positive feedback architecture}
\label{sec:lacoperon}

\begin{figure}[h!]
\begin{center}
\includegraphics[width=0.95\textwidth]{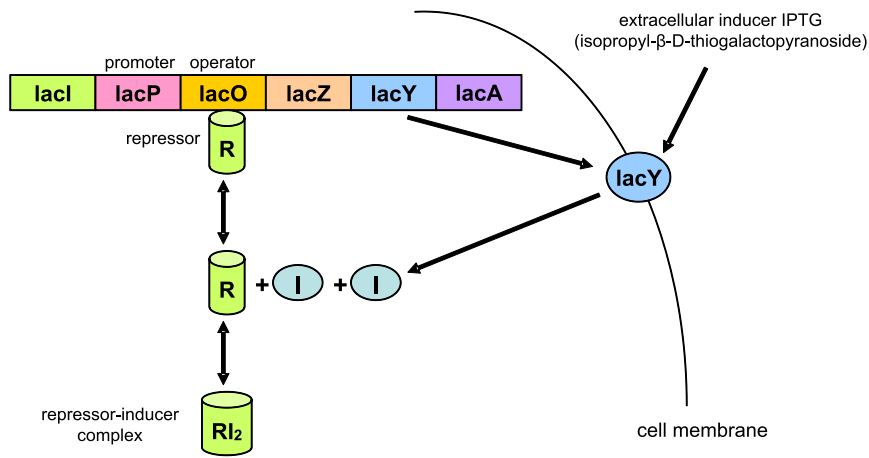}
\end{center}
\caption{Sketch of the positive feedback loop network, {\it lac} operon with IPTG induction}
\label{fig:lacoperon}
\end{figure}

The {\it lac} operon consists of the promoter {\it lac}P, the operator {\it lac}O and three genes, which encode the proteins required for the metabolism of lactose (see Fig.~\ref{fig:lacoperon}).
The gene {\it lac}Z encodes the enzyme $\beta$-{\it galactosidase} and the {\it lac}A encodes transacetylase.
The gene {\it lac}Y is responsible for the encoding of the protein {\it lac} permease, which is the potent transport facilitator.
{\it Lac} permease contributes to the transport of lactose or an analogue, such as TMG or IPTG in the interior of the cell. 
In the absence of lactose or of an extracellular inducer (TMG, IPTG), the expression of {\it lac} operon genes is turned off by the inhibitive action of {\it lac}I.
%

The inhibitor {\it lac}I binds to the operator site, {\it lac}O, prevents
binding of the RNA polymerase thus inhibiting the transcription of the
genes' DNA into the corresponding mRNA.
On the other hand, in the presence of lactose, TMG or IPTG, the inducer is transported into the cell, binds to the repressor {\it lac}I through a bimolecular reaction and the operator {\it lac}O becomes free of {\it lac}I, hence initiating the transcription.
Upon expression of {\it lac}Y, further transport of the inducer occurs at a higher rate resulting in further expression of the three {\it lac} operon genes.
Thus, the expression of {\it lac}Y gene promotes its further expression, and the network functions as an autocatalytic system or a positive feedback loop.
Genetic networks with genes carrying the positive feedback architecture exhibit bistable behavior at the single-cell level \cite{Mantzaris:2007:FSP,Kavousanakis:2009}.
A simple mathematical model, which captures the basic features of the positive feedback loop architecture is described in Mantzaris (2005)..
A simplified description of the reaction steps has been proposed by Kepler and Elston (2001):

\begin{eqnarray}
O_0 &\overset{k_0}{\rightarrow} Y \\
O_1 &\overset{k_1}{\rightarrow} Y \\
Y &\overset{\lambda}{\rightarrow} \varnothing \\
O_0+Z & \overset{\phi}{ \underset{\alpha \phi}{\rightleftarrows}} O_1 \\ 
Y+Y & \overset{\chi}{\underset{\beta \chi}{\rightleftarrows}} Z.
\end{eqnarray}
\\

The $O_0$, $O_1$ symbols denote the fraction of free and occupied operator sites;
$Y$ is the monomer produced by the expression of {\it lac}Y gene in either the occupied or the unoccupied state of the operator {\it lac}O.
The rate of {\it lac}Y expression in the unoccupied state ($k_0$) is significantly lower than that in the occupied state ($k_1$).
$Z$ is the dimer of the gene product, which binds to the free operator site leading to an occupied state.
Assuming that the production rates of the monomer product are proportional to the fractions of unoccupied and occupied operator sites and that the degradation of $Y$ is a first order reaction, the single-cell monomer dynamics are described by:

\begin{equation}
\label{eq:monomer}
\frac{\textrm{d}Y}{\textrm{d}\tau}=k_0 O_0 + k_1 O_1 -\lambda Y,
\end{equation}

\noindent where $\lambda$ is the degradation rate constant.
The number of operation sites is conserved:

\begin{equation}
\label{eq:conserv}
O_0 + O_1 =1.
\end{equation}

It is further assumed that the unoccupied and occupied states are in equilibrium and the same holds for the dimerization reaction:

\begin{eqnarray}
\label{eq:dimerization}
O_0 Z = \alpha O_1, \\
Y^2 = \beta Z,
\end{eqnarray}

\noindent where $\alpha$ and $\beta$ are the equilibrium constants of the operator transition and dimerization reactions, respectively.
Substituting (\ref{eq:conserv}) and (\ref{eq:dimerization}) into (\ref{eq:monomer}) yields the following expression for the reaction rate:

\begin{equation}
\label{eq:finalmonomer}
\frac{\textrm{d}Y}{\textrm{d}\tau}=\frac{ k_0 \alpha \beta +k_1 Y^2}{\alpha \beta +Y^2} -\lambda Y.
\end{equation}

In order to obtain the dimensionless form of the reaction rate $R(x)$, we non-dimensionalize the intracellular content $Y$ and time $\tau$ as follows:

\begin{equation}
x=\frac{Y}{Y^*} \quad \tau = \frac{t}{t*}.
\end{equation}

\begin{figure}[h!]
\begin{center}
  \includegraphics[width=0.75\textwidth]{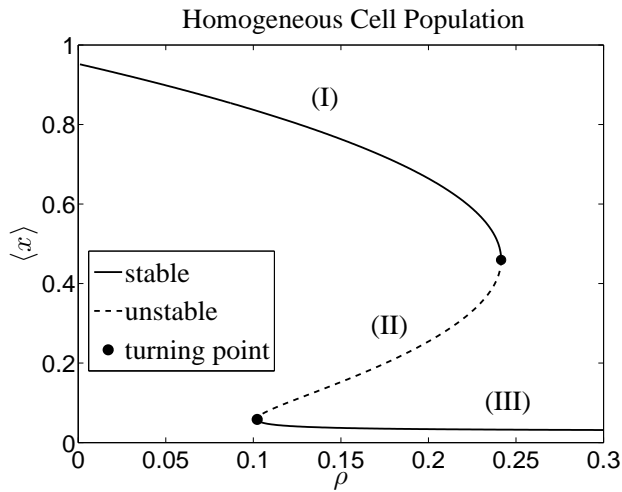}
  \end{center}
\caption{Dependence of average intracellular {\it lac}Y content, $\left< x \right>$, of a homogeneous cell population on the inverse IPTG inducer concentration, $\rho$.
Parameter set values: $\pi=0.03$, and $\delta=0.05$.}
\label{fig:single_bistable}       % Give a unique label
\end{figure}

Setting: $k_1t^*/y^*=1$, $\pi = k_o/k_1$, $\rho=\alpha \beta / {y^*}^2$, $\delta = \lambda t^*$ \cite{Mantzaris:2005:CPB}, and substituting to (\ref{eq:finalmonomer}) the following non-dimensional form for the rate of change of the dimensionless {\it lac}Y amount (the intracellular content $x$) is obtained:

\begin{equation}
\label{eq:lacY}
\frac{\textrm{d}x}{\textrm{d}t} \equiv R(x) = \frac{ \pi \rho  + x^2}{\rho +x^2} -\delta x,
\end{equation}

\noindent where $\pi$ is the relative rate of expression when the operator {\it lac}O is either free or occupied ($\pi << 1$, due to significantly lower rate of gene expression while being in the unoccupied state);
$\rho$ is a parameter inversely proportional to the extracellular inducer concentration (IPTG) and $\delta$ is the dimensionless degradation rate.
The main feature of {\it lac} operon is that its dynamics exhibit a bistable behaviour, where two stable steady stats co-exist with an unstable steady state within a significant region of the parameter space $(\pi, \rho, \delta)$.
If we examine the simplest case of a cell population, where all individuals behave like the average cell (homogeneous population), then the number cell density is expressed as $n(x,t)=\delta \left( x- \left< x \right> \right)$, where $\delta$ is the Dirac function.
By introducing the number cell density expression for the homogeneous cell population to (\ref{eq:average}), then the dynamics of $\left< x \right>$ are described from the following ordinary differential equation:
\begin{equation}
\label{eq:homogeneous}
\frac{\textrm{d} \left< x \right>}{\textrm{d} t} = R \left( \left< x \right> \right) - \left< x \right> = \frac{ \pi \rho + \left< x \right>^2}{ \rho+ \left< x \right>^2} -\delta \left< x \right> -\left< x \right>.
\end{equation}
In Fig.~\ref{fig:single_bistable}, we present the dependence of the average intracellular content of a homogeneous cell population, $\left<x \right>$, as a function of the inverse proportional of the IPTG concentration, $\rho$.
The upper branch (I) corresponds to stable steady state solutions with high expression levels of {\it lac}Y, and the lower branch (III) to solutions with low expression level of {\it lac}Y. 
The intermediate branch of solutions corresponds to dynamically unstable steady states, i.e., an infinitesimally small perturbation drives the system towards one of the co-existing stable steady state solutions (branch (I) or (III)). 
There exists a wide parametric region, $\rho \in [0.102, 0.242]$, within which the cell population exhibits bistable behaviour and more than one asymptotic phenotypes can be observed.

The naturally arisen question is whether this bistable behaviour persists in heterogeneous cell populations, and what is the extend of bistability when heterogeneity comes into play. 
In Kavousanakis et al. (2009) it is demonstrated through the numerical solution of the deterministic CPB model, that indeed cell heterogeneity has a significant impact on the range of bistability shifting this area towards higher IPTG values.
As mentioned above, the CPB model is suitable for large size populations, where the continuum assumption can be justified.
In order to study the effect of heterogeneity on small size cell populations, in which random events during division is of significant importance, the most appropriate approach is through stochastic modelling.
Here, we apply the CNMC model \cite{Mantzaris:2006:SDS} for a cell population with the {\it lac} operon network and present direct temporal simulation results in the next paragraph.

\subsection{Direct temporal simulations of the stochastic CNMC algorithm}
\label{sec:direct_temporal}
\begin{figure}[h!]
\begin{center}
  \includegraphics[width=0.95\textwidth]{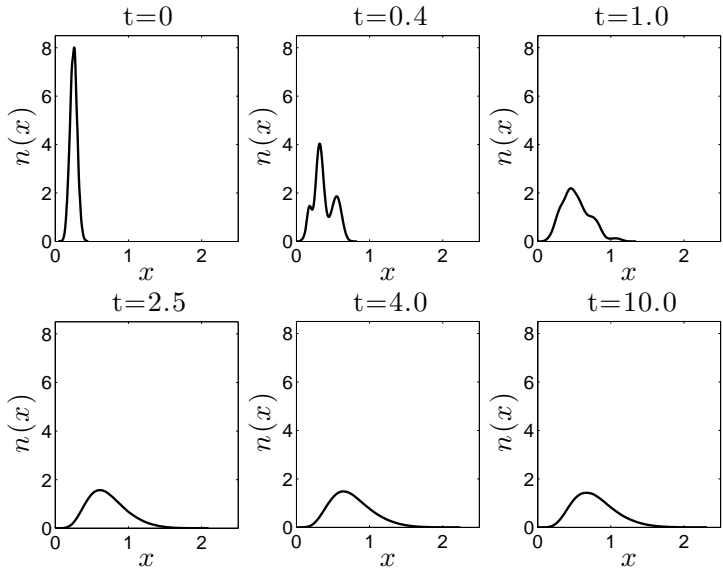}
  \end{center}
\caption{Snapshots of the time evolution of the density function of a population of 10,000 cells simulated with the CNMC model.
Parameter set values: $f=0.5$, $m=2$, $\pi=0.03$, $\delta=0.05$, and $\rho=0.06$.}
\label{fig:stochastic_evolution}       % Give a unique label
\end{figure}

In Fig.~\ref{fig:stochastic_evolution} we present snapshots of the evolution of the density function of a large size population (10,000 cells), carrying the {\it lac} operon genetic network.
The initial condition is a shifted Gaussian distribution with mean value $\mu=0.25$ and standard deviation $\sigma=0.05$ ($N(0.25,0.05^2)$).
The density function $n(x,\tau)$ denotes the number of cells of dimensionless content $x$ ({\it lac}Y) at dimensionless time $\tau$, divided by the total number of cells. 
The results are produced by incorporating: (a) Eq. (\ref{eq:lacY}), which describes the {\it lac}Y rate of change, (b) Eq. (\ref{eq:division}) which describes the division rate of each cell, and (c) the discrete partitioning mechanism (\ref{eq:partition}) with $f=0.5$, in the CNMC model (described in Sec.~\ref{sec:kmcmodel}).
During the initial stage of the simulation the Gaussian distribution splits into a three-humped distribution ($\tau=0.4$), and gradually evolves towards a smoother single-humped distribution, which remains practically invariant after $\tau=2.5$ (steady state).

\begin{figure}[h!]
\begin{center}
  \includegraphics[width=0.75\textwidth]{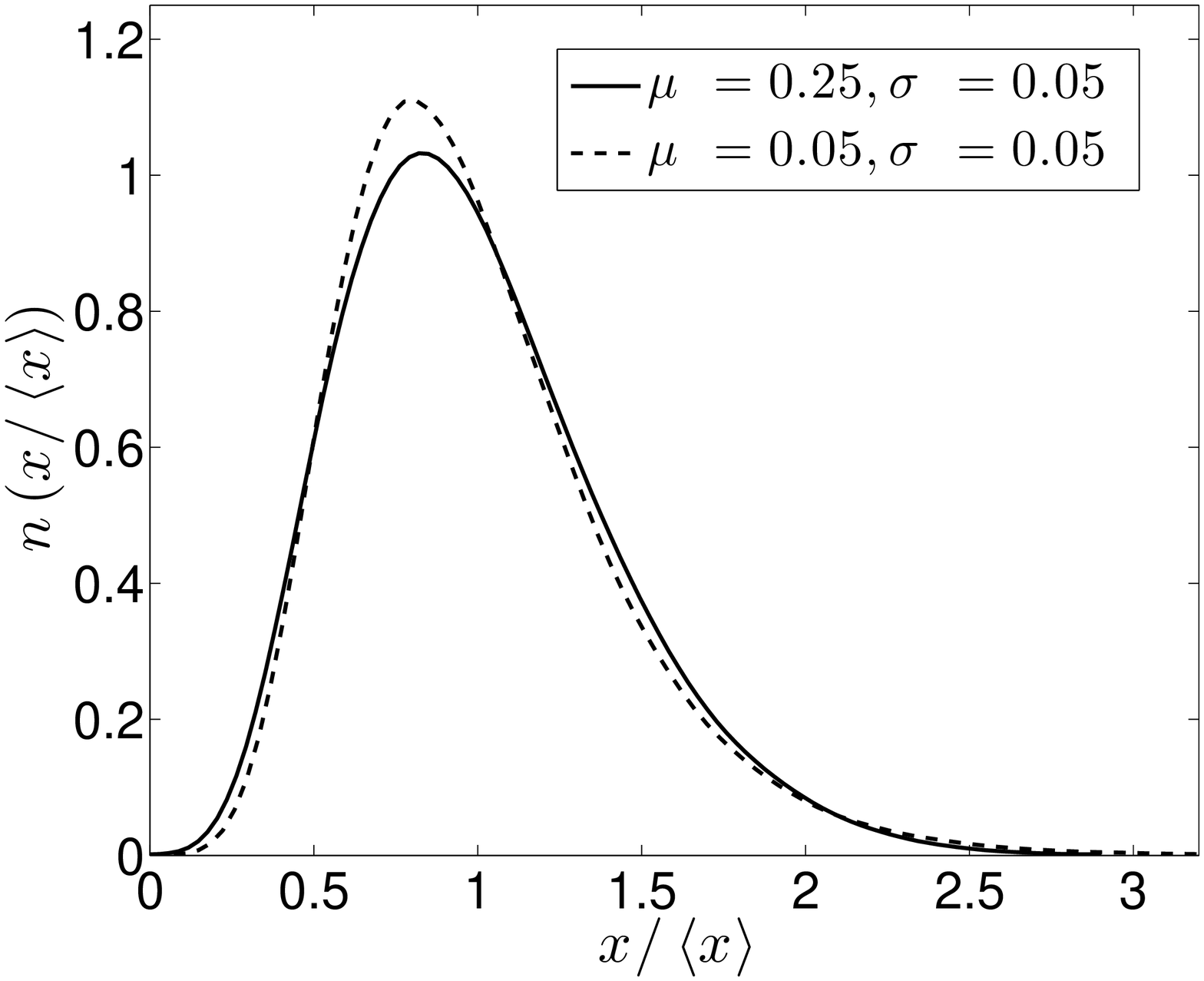}
  \end{center}
\caption{Asymptotic states (average of 100 copies for noise reduction purposes) of a population of 10,000 cells simulated with the CNMC model, and different initial conditions.
Both initial conditions are shifted Gaussian distributions with mean $\mu$ and standard deviation $\sigma=0.05$.
The initial condition with $\mu=0.25$ converges to the steady state solution depicted with the dashed line ($\left< x \right>=0.61$) and the initial state with $\mu=0.05$ to the steady state distribution depicted with the solid line ($\left< x \right>=0.04$).
Parameter set values: $f=0.5$, $m=2$, $\pi=0.03$, $\delta=0.05$, and $\rho=0.10$.}
\label{fig:asymptotic_coexist}       % Give a unique label
\end{figure}

As mentioned above, it is expected that the isogenic cell population exhibits solution multiplicity within a range of the parameter $\rho$.
In order to compute the possibly co-existing steady state solutions through direct temporal simulations, one needs to choose appropriate initial conditions. 
Indicatively, we present in Fig.~\ref{fig:asymptotic_coexist} two asymptotic (steady) states (normalised with respect to their average expression $\left<x\right>$) for the same set of parameter values: $f=0.5$, $m=2$, $\pi=0.03$, $\delta=0.05$, and $\rho=0.10$, and different initial states.
The solid line distribution has an average intracellular content of $\left<x\right>=0.61$, when the initial condition is a shifted Gaussian distribution $N(\mu=0.25, \sigma^2=0.05^2)$. 
The asymptotic distribution depicted by the dashed line, has an average intracellular content of $\left<x\right>=0.04$, when the initial condition is a shifted Gaussian distribution with a smaller mean value of $\mu=0.05$ and the same standard deviation, $\sigma=0.05$. 
Thus, solution multiplicity is also predicted by performing stochastic simulations of a cell population.
However, the approach of exploring the solution space of steady states over a wide range of parameter values through direct temporal simulations is not computationally efficient.
The computational load is significantly high, since it is required to perform long time interval executions of direct temporal simulations for different sets of parameter values.
Furthermore, the accurate determination of the bistability region is not guaranteed (if not impossible), since stochastic noise at the vicinity of the bistability region (turning points) can cause unpredictable phenotypic switches. 

In order to perform such an analysis for a system the description of which is available at a fine-microscopic level, it is required to resort to alternative methods of multiscale analysis.
In this work, we adopt ideas from the equation-free methodology framework \cite{Kevrekidis:2003}, which enables the performance of coarse-level numerical tasks, utilising information which originates from short bursts of microscopic simulations. 
A description of this methodology and its application to the studied system of the isogenic cell population is provided in the next section.

\section{Coarse-grained analysis}
\label{sec:coarsemethod}

We are interested in studying the macroscopic behaviour of the heterogeneous cell population, utilising simulations which are performed at a microscopic / cell-level.
In particular, our target is to compute the coarse asymptotic behaviour of cell populations phenotype carrying the {\it lac} operon genetic network , and explore its dependence on the extracellular inducer concentration (IPTG). 
When closed descriptions of the macroscopic variables under study are available, the performance of systems-level tasks, such as the computation of steady state solutions, and bifurcation analysis is feasible through an arsenal of analytical and numerical techniques.
In our case study, such deterministic descriptions are available (CPB models), however their validity is limited only when large size populations are simulated.
In order to study the steady state behaviour of smaller size populations, one would need to perform an extensive number of direct stochastic simulations, for adequately large time intervals, so as to ensure the convergence of the system to a ``coarse'' steady state solution.
However, the computational requirements for such an approach are prohibitively large;
in addition, and since the studied system of the {\it lac} operon network exhibits bistability, noise effects can lead to abrupt phenotypic changes when the parameter value is at the proximity of critical turning points. 
Here, we adopt a computational framework, which enables the performance of systems-level tasks, by utilising short bursts of appropriately initialised stochastic simulations (instead of performing long time executions).
The backbone of this framework, commonly known as the ``equation-free'' approach \cite{Gear:2002,Siettos:2003,Kevrekidis:2004}, is the ``coarse time-stepper''.

\subsection{Coarse time-stepper}
\label{sec:coarse_ts}

\begin{figure}[h!]
\begin{center}
  \includegraphics[width=0.95\textwidth]{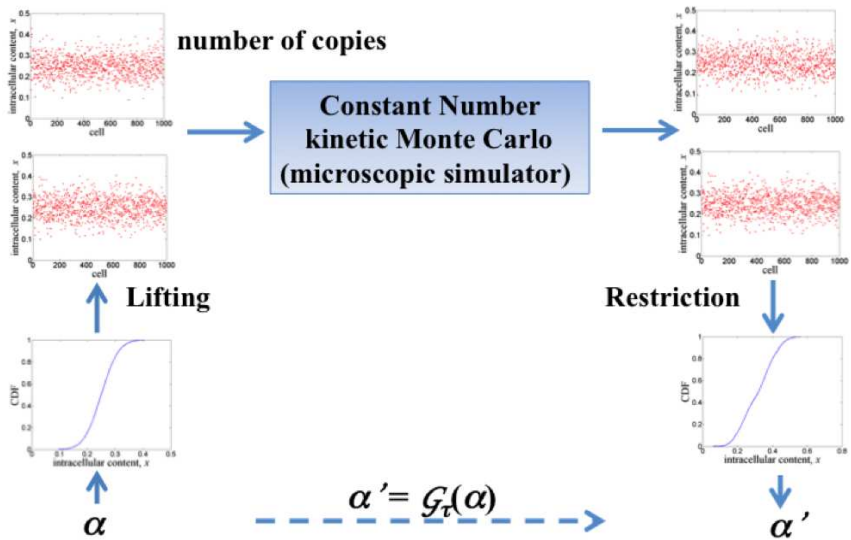}
  \end{center}
\caption{Schematic of the coarse time-stepper for the model of an isogenic cell population carrying the {\it lac} operon network.}
\label{fig:coarse_ts}       % Give a unique label
\end{figure}

Suppose we are interested in studying the dynamics and/or asymptotic behaviour of a coarse variable $f$. 
In our case, $f$ is taken to be the cumulative distribution function of cells, which are simulated by the CNMC model.
An upscaling of the discrete model should yield explicit equations governing the dynamics of $f$, e.g., the deterministic CPB models, which are valid under strong assumptions (continuum assumption).
An alternative approach, which can bypass the derivation of such closures is to apply the so-called {\it coarse time-stepper}, a schematic of which is shown in Fig.~\ref{fig:coarse_ts}.
The coarse time-stepper employs two steps that link the fine-scale simulator with coarse-level computations: {\it restriction} and {\it lifting}

\noindent{\it Restriction}
\noindent Suppose that the stochastic simulation of $N$ cells yields a set of intracellular {\it lac}Y contents: ${\bf x}=\{x_i\}, i=1,..,N$.
Then the cumulative distribution function (CDF) $f$ is trivially computed  by sorting into an ascending order the $N$-dimensional ${\bf x}$ vector
($x_i \leq x_{i+1}$) and plot the graph, which consists of the points: $({\bf x}, {\bf p}) = \left( x_i, p_i=(i-0.5)/N \right), i=1,...,N$.
This provides our restriction of microscopic data, $\bf U$, to the macroscopic description, ${\bf f }$, through an operator $\mathcal{M}$, i.e., ${\bf f} = \mathcal{M} {\bf U} $.
In case of many repetitions of the same stochastic simulation, we compute the average restriction. 
\newline

\begin{figure}[h!]
\begin{center}
\begin{tabular}{c c}
  \includegraphics[width=0.5\textwidth]{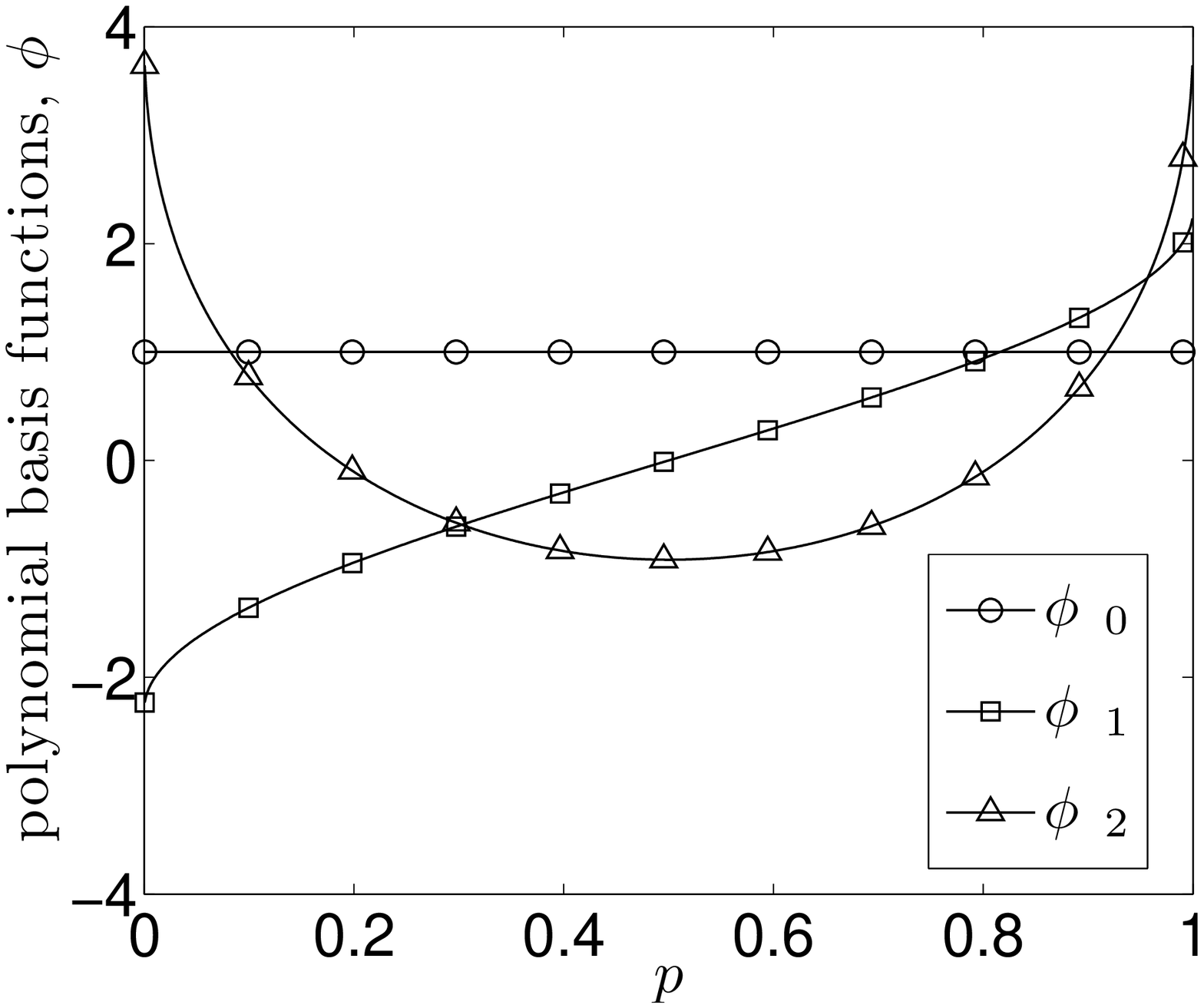} & \includegraphics[width=0.5\textwidth]{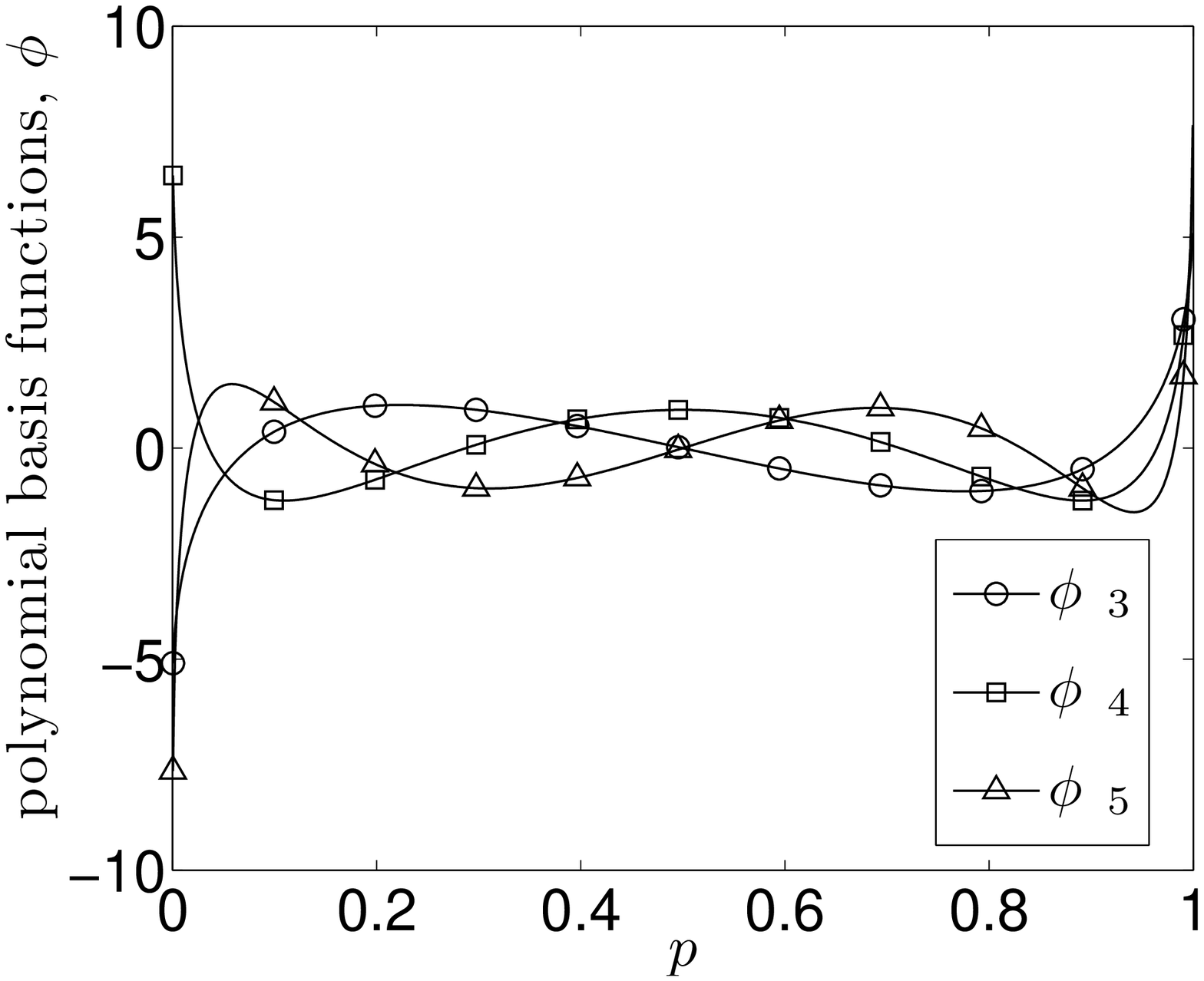} \\
  (a) & (b)
  \end{tabular}
  \end{center}
\caption{Graphical representation of the orthogonal polynomial basis functions (up to $q=5$ order) for the coarse description of cell populations.}
\label{fig:basis}       % Give a unique label
\end{figure}

\noindent{\it Lifting}
\noindent We choose to work with the CDF rather than the -natural to think- probability density function (PDF), $n$, because it is the derivative of a numerically specified function (the CDF) and numerical differentiation is poorly conditioned. 
In practice, it is easier to work with the inverse CDF (ICDF), ${\bf x}({\bf p})$, which gives the intracellular content $x_i = x(p_i)$ 
of a given cell, $i$.
Assuming, that the ICDF is smooth enough, we can use a low-dimensional description of it based on the first few of an appropriate sequence of orthogonal polynomials \cite{Gear:2001}.
Orthogonal polynomials are simple approximations, yet they are not useful for the CDF which has a possible support from $- \infty$ to $+ \infty$.
However, the inverse CDF, $x(p)$ has a finite support ($p \in [0,1]$) and can be easily approximated by a series of $j$-degree orthogonal polynomials $\phi_j(p)$, which must be monotone, non-decreasing and must lie between 0 and 1:
\begin{equation}
\label{eq:orthog_polyn}
x(p) \approx \sum_{j=0}^q \alpha_j \phi_j (p).
\end{equation}
Computationally, the $\phi_j, j=0,...,q$ polynomials are represented by their values on the point set ${\bf p}$ and
it is seldom to find it necessary to use more than the first six basis functions ($q=5$). 
In Fig.~\ref{fig:basis}(a)-(b), we present the up to $5^{th}$ order degree basis functions, which are used for the coarse description of the heterogeneous cell population.
They can be pre-computed (they depend only on $N$) and form a $(q+1) \times N$ matrix, $\Phi$.
Then the coefficients $\alpha = {\alpha_i}$, $i=0,...,q$ in Eq.~(\ref{eq:orthog_polyn}) are computed from $\alpha = \Phi {\bf x}$, and the approximation 
to ${\bf x}$ can be computed from ${\bf x} \approx \Phi^T \alpha$.
Thus, we can evaluate the intracellular content of each cell using a small set of $\alpha$ values.
This constitutes the lifting step, which maps the macroscopic data to consistent microscopic realisations through an operator, $\mu$, i.e., ${\bf U} = \mu {\bf f} $.
This choice of lifting and restriction operators ensures that lifting from macroscopic to the microscopic and then restricting down again has no effect (except round-off), i.e., $\mu \mathcal{M} \approx I$.

Given the vector of $\alpha$ values we generate through lifting a consistent microscopic realisation (intracellular content values $x$ for $N$ cells).
We perform a number of CNMC simulations for the $N$-cell population and after a short time interval $\tau$ we obtain its restriction (average CDF, and updated $\alpha$ values).
Through lifting and restriction steps, we sidestep the derivation of a closure for the evolution of the macroscopic $\alpha$ variables;
instead we construct a discrete time-mapping:

\begin{equation}
\label{eq:time_horizon}
\alpha' = \mathcal{G}_\tau (\alpha),
\end{equation}
 
\noindent which can be utilised for the performance of black-box based numerical tasks.
One of these tasks is the computation of the coarse steady state  $\alpha^*$, by solving the following set of non-linear equations:
\begin{equation}
{\bf R} \equiv \alpha^* - \mathcal{G}_\tau (\alpha^*) = {\bf 0},
\end{equation}

\noindent with the Newton-Raphson method, which requires during each iteration the solution of the linearised system:
\begin{eqnarray}
\frac{ \partial {\bf R}}{\partial \alpha} \delta \alpha = - {\bf R} \rightarrow \nonumber \\
\left[ {\bf I} - \frac{\partial \mathcal{G}_\tau}{\partial \alpha} \right] \delta \alpha = -{\bf R}.
\end{eqnarray}

\noindent We re-iterate that an explicit expression of the operator $\mathcal{G}_\tau$ is not available;
however, we can still approximate it numerically by appropriate initialisation of the coarse time-stepper.
For example, the $(i,j)$ element of matrix $\frac{\partial \mathcal{G}_\tau (\alpha) }{\partial{\alpha}}$ is approximated by the forward finite difference scheme:
\begin{equation}
\left( \frac{\partial \mathcal{G}_\tau}{\partial{\alpha}} \right)_{i,j} \approx \frac{ \left(\mathcal{G}_{\tau} (\alpha_j + \epsilon) \right)_i - \left( \mathcal{G}_{\tau} (\alpha_j) \right)_i}{\epsilon},
\end{equation}

\noindent where $\epsilon$ is a small number (perturbation number).
Furthermore, an extension of this application is the performance of bifurcation analysis by means of pseudo-arc-length parameter continuation techniques \cite{Keller:1977}.
Such techniques enable the computation of both stable and unstable coarse steady state solutions, and the accurate computation of bistability limits. 
In the computations presented below, the continuation parameter is the (dimensionless) inverse concentration of the extracellular inducer IPTG, $\rho$.
In addition, the stability of each computed solution can be quantified by determining the eigenvalues of the matrix,  $\frac{\partial \mathcal{G_\tau \left(\alpha^*\right)}}{\partial{\alpha}}$.
When at least one eigenvalue crosses the unit circle in the complex plane, the solution is characterised as dynamically unstable \cite{Strogatz:1994}.

\section{Coarse-grained parametric analysis of heterogeneous cell population}
\label{sec:results}

\begin{figure}[h!]
\begin{center}
  \includegraphics[width=0.65\textwidth]{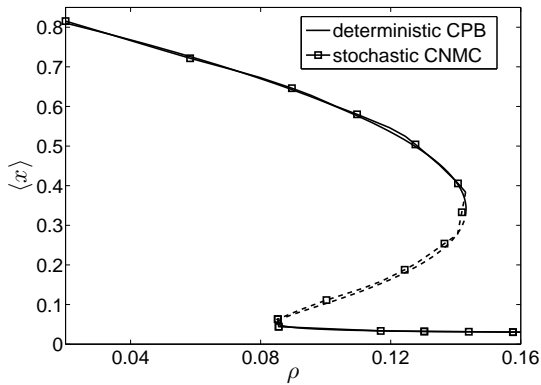}
  \end{center}
\caption{Dependence of steady state average expression level, $\left<x \right>$ on the inverse IPTG concentration $\rho$.
The black lines (solid and dashed) correspond to results obtained from the deterministic CPB model, whereas the lines with open squares correspond to the stochastic CNMC model executed with 10,000 cells (average of 100 copies for noise reduction).
The dashed lines in both the deterministic and stochastic models correspond to unstable steady state solutions.
Parameter set values: $f=0.5$, $m=2$, $\pi=0.03$, and $\delta=0.05$.}
\label{fig:validate}       % Give a unique label
\end{figure}

\begin{figure}[h!]
\begin{center}
\begin{tabular}{cc}
  \includegraphics[width=0.5\textwidth]{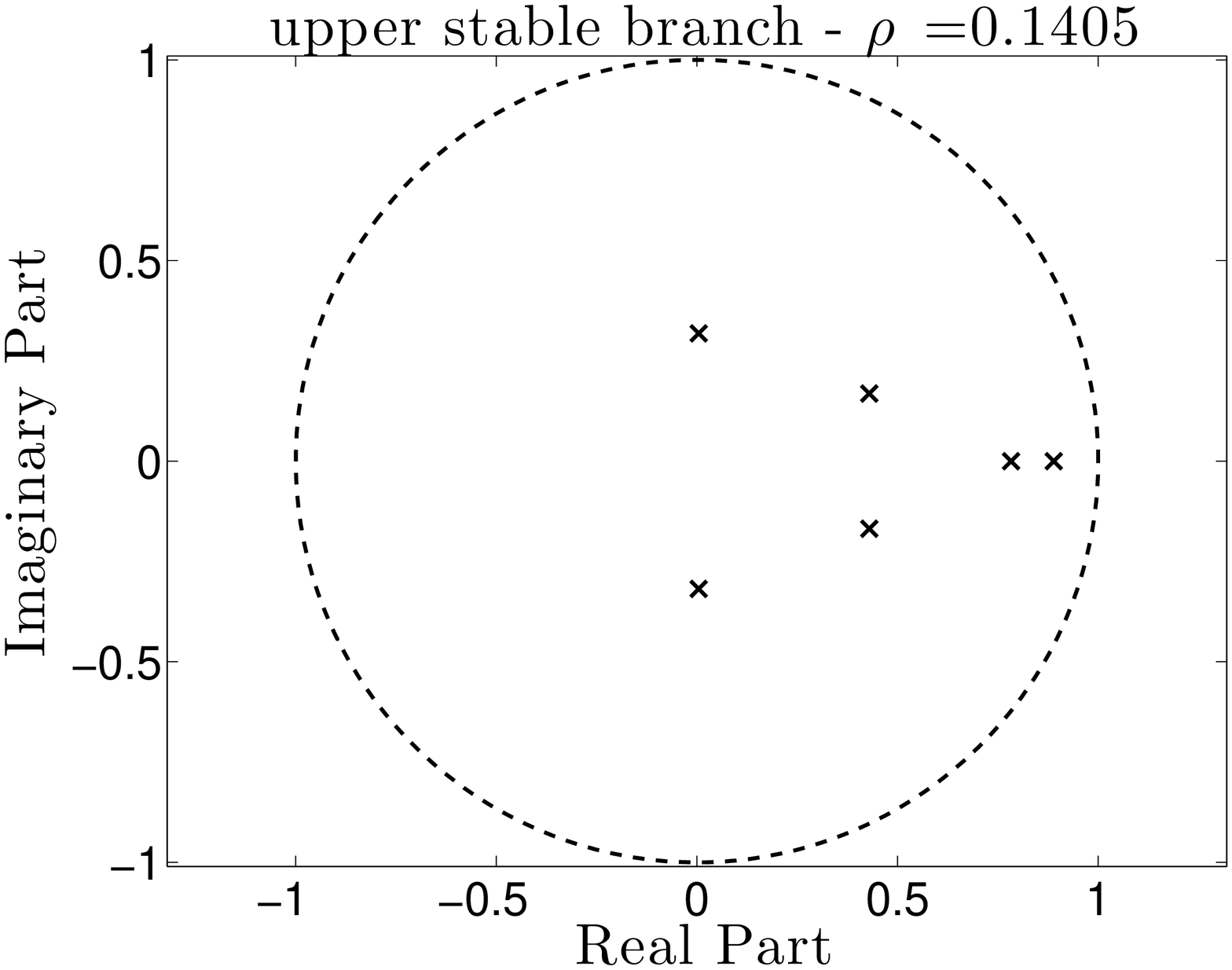} &
  \includegraphics[width=0.5\textwidth]{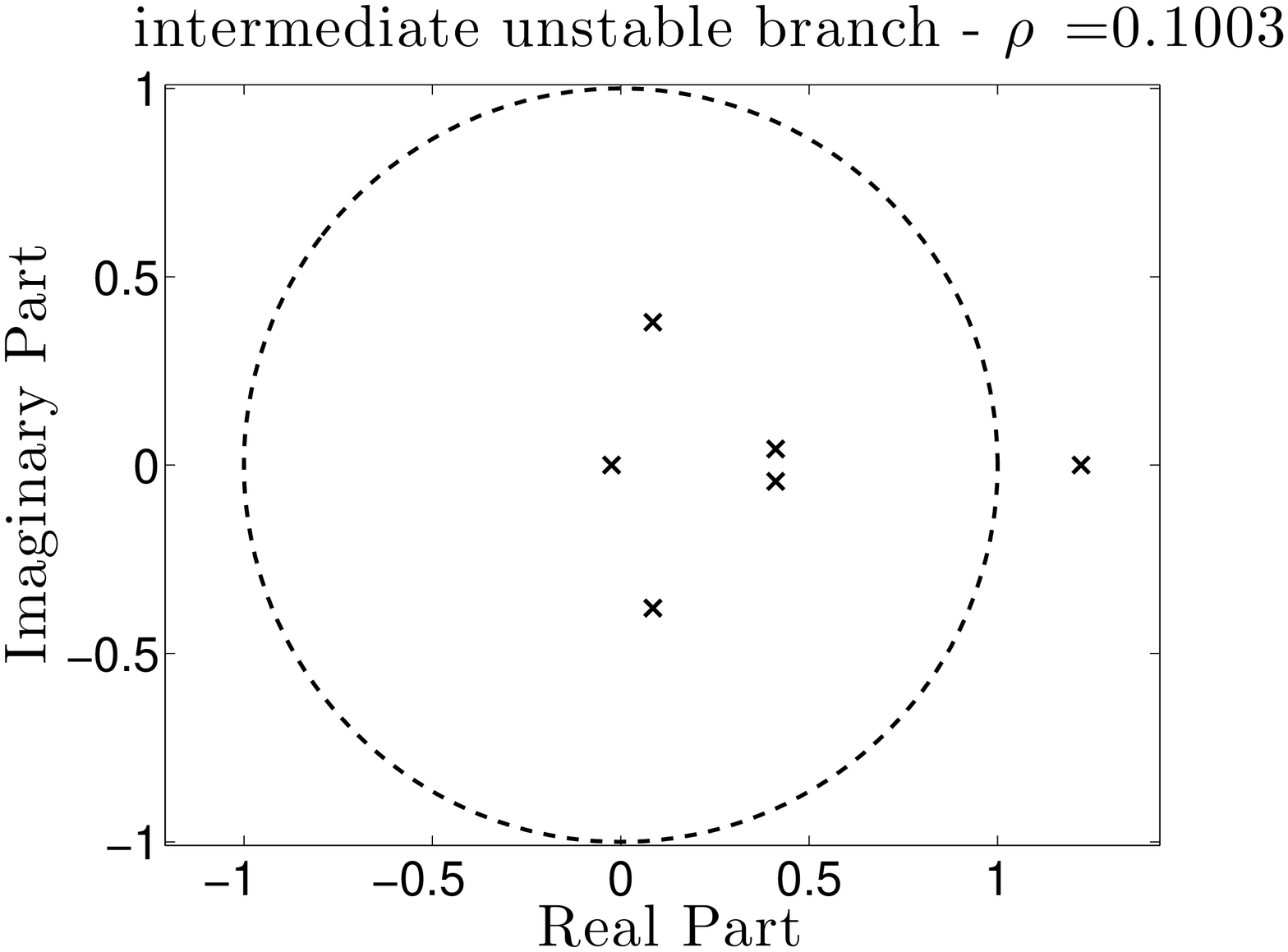} \\
  (a) & (b) 
  \end{tabular}
  \begin{tabular}{c}
  \includegraphics[width=0.5\textwidth]{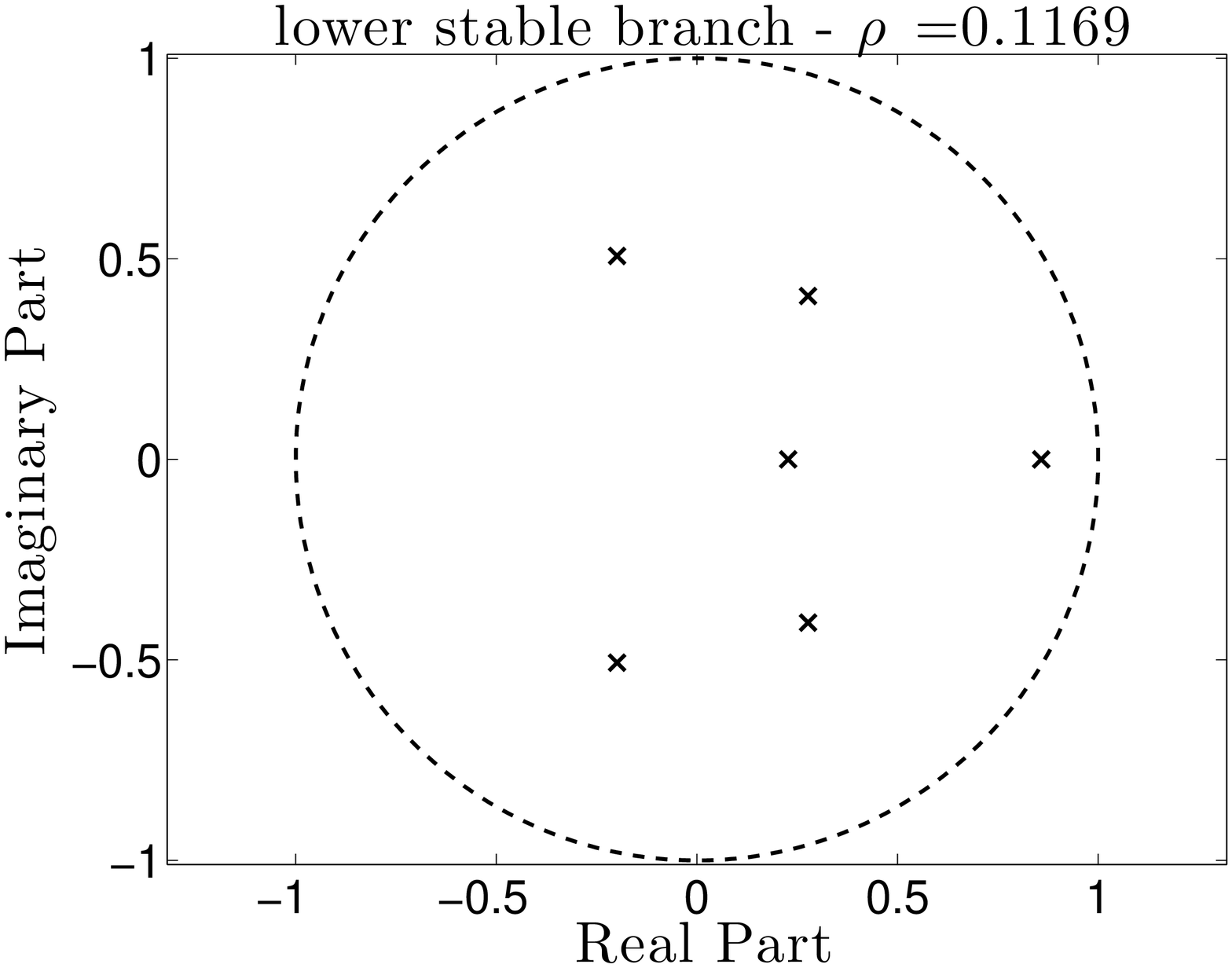} \\
  (c)
  \end{tabular}
 \end{center}
\caption{Results of coarse stability analysis.
Eigenvalues of the $\frac{\partial \mathcal{G_\tau \left(\alpha^*\right)}}{\partial{\alpha}}$ matrix corresponding to (a) an upper branch stable steady state ($\rho \approx 0.14$), (b) an intermediate branch unstable steady state ($\rho \approx 0.1$), and (c) a lower branch steady state solution ($\rho \approx 0.12$).
The loss of stability in (b) is marked by the eigenvalue crossing the unit circle (dashed line) in the complex plane.
Parameter set values: $f=0.5$, $m=2$, $\pi=0.03$, and $\delta=0.05$,$N=10,000$ cells.}
\label{fig:spectra}       % Give a unique label
\end{figure}

\begin{figure}[h!]
\begin{center}
\begin{tabular}{c}
  \includegraphics[width=0.65\textwidth]{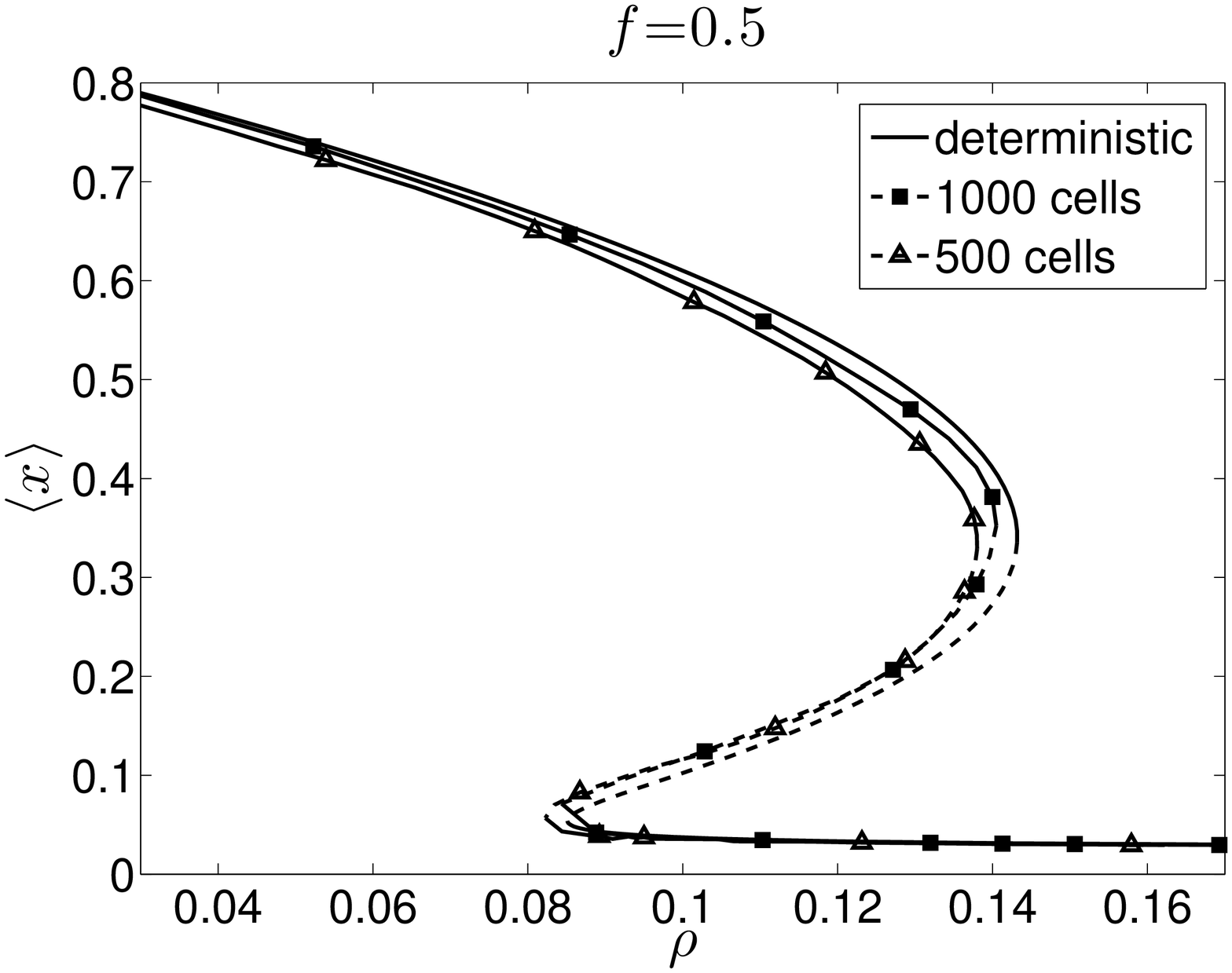} \\
  (a) \\
  \includegraphics[width=0.65\textwidth]{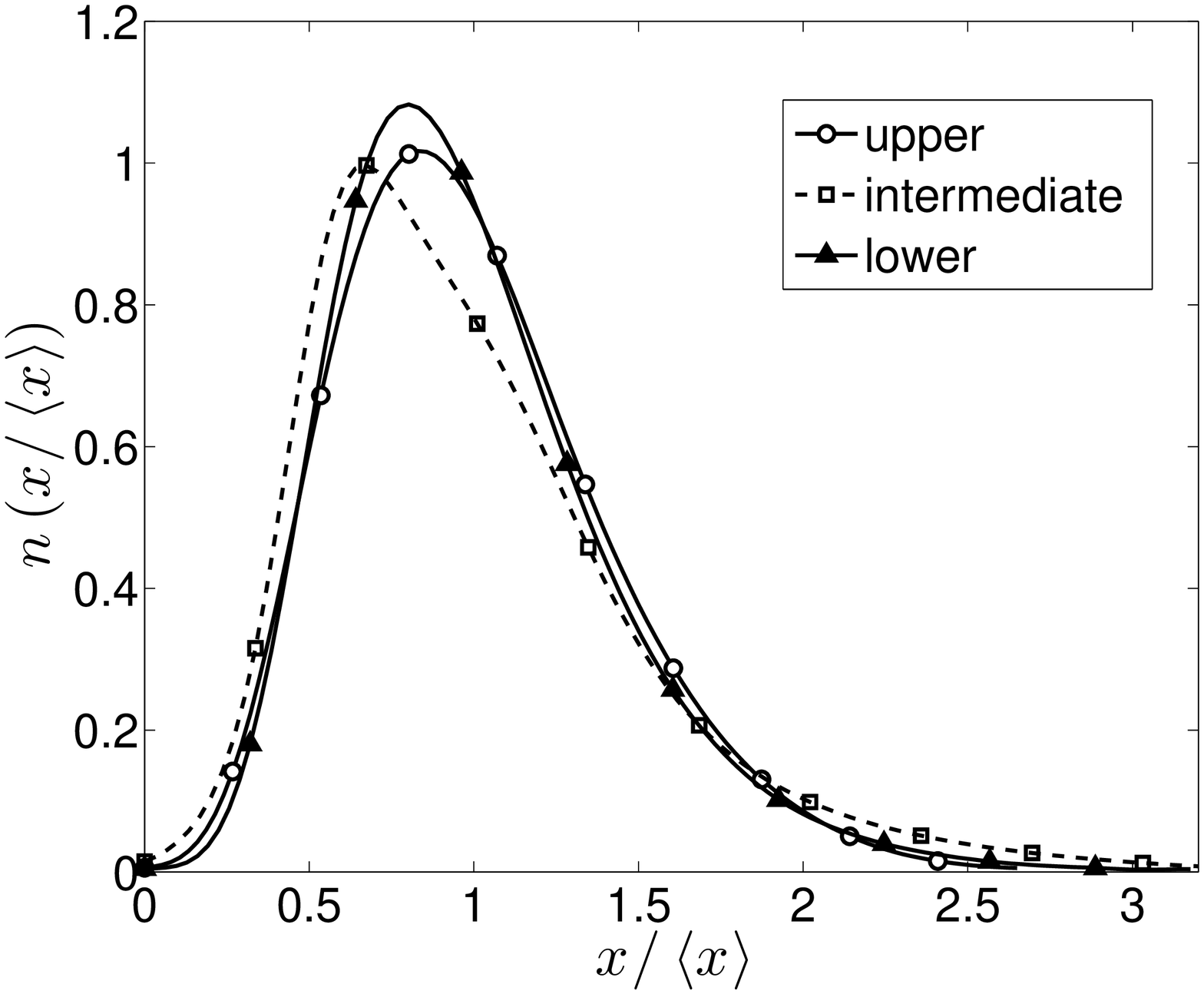} \\
  (b)
  \end{tabular}
  \end{center}
\caption{(a) Effect of the population size on the average
expression of the {\it lac}Y gene steady state ($\left<x \right>$), as a function of
the inverse IPTG concentration $\rho$.
The black line corresponds to the deterministic CPB model, the line with full squares to corresponds to CNMC simulation with $N=1000$ cells,
and the line with open triangles corresponds to the CNMC simulation with $N=500$ cells.
Solid and dashed lines correspond to stable and unstable steady states in all cases.
(b) Three coexisting (normalised with respect to $\left< x \right>$) coarse steady state solutions at $\rho=0.0963$ obtained from CNMC simulations with $N=1000$ cells (average of 1000 copies).
The solid line with open circles corresponds to the stable upper branch solution, the dashed line with open rectangles corresponds to the unstable intermediate branch solution, and the solid line with full triangles depicts the lower stable branch solution.
Parameter set values: $f=0.5$, $m=2$, $\pi=0.03$, and $\delta=0.05$.}
\label{fig:m_eq_2_f_eq_05}       % Give a unique label
\end{figure}

In this section, we apply the equation-free methodology in order to perform a parametric analysis of the asymptotic behaviour of heterogeneous cell population as a function of the extrcellular inducer IPTG.
In the results presented below, we choose as continuation parameter the inverse of the extracellular IPTG inducer ($\rho$), which can be adjusted experimentally. 
First, we validate our computational methodology by comparing the findings of coarse bifurcation analysis when applied on a large size cell population against results obtained from the numerical solution of the deterministic CPB model. 
In particular, we examine the case of a 10,000 cell population with symmetric partitioning mechanism ($f=0.5$).
In Fig.~\ref{fig:validate}, we present a typical bifurcation diagram, showing the steady state behaviour of the average intracellular {\it lac}Y content as a function of the dimensionless $\rho$ parameter. 
The comparison between the results obtained from the stochastic CNMC model and the deterministic CPB model shows very good agreement, even when predicting unstable steady state solutions (dashed lines in Fig.~\ref{fig:validate}).  
The computation of both stable and unstable steady state solution is feasible through the application of pseudo arc-length parameter continuation techniques, performed within the equation-free framework. 
This application enables the tracing of the entire solution space, contrary to the performance of solely long time interval stochastic simulations, which require appropriately chosen initial conditions in order to converge to the upper or lower branch of stable steady state solutions (see also discussion in Sec.~\ref{sec:direct_temporal}).
Furthermore, we can quantify the stability of the obtained steady state solutions be determining the eigenvalues of the $\frac{\partial \mathcal{G_\tau \left(\alpha^*\right)}}{\partial{\alpha}}$ matrix.
In Fig.~\ref{fig:spectra} we present the results of coarse stability analysis as obtained for (a) an upper branch (b) intermediate branch and (c) lower branch steady solution. 
In Fig.~\ref{fig:spectra}(a) and (c), all eigenvalues lie within the limits of the complex plane unit circle, and the corresponding steady state solutions are characterised as stable. 
The existence of one eigenvalue crossing the complex plane unit circle in Fig.~\ref{fig:spectra}(b), renders the corresponding steady state solution as dynamically unstable.
Upon validation of our methodology, we can now investigate the effect of noise effects during division on the range of bistability, which is increasing for small size populations. 
In particular, we perform coarse bifurcation analysis for cell population sizes of $N=500$ and $N=1000$ cells and compare the results with the deterministic CPB model (\ref{eq:determ}), which is numerically solved with the finite element method \cite{Kavousanakis:2009}.
The bifurcation diagram of the steady state expression level of the average intracellular content $\left<x \right>$ as a function of the dimensionless parameter $\rho$ is presented in Fig.~\ref{fig:m_eq_2_f_eq_05}(a), when equal partitioning is considered ($f=0.5$).
One can observe the shift of the right turning point towards lower $\rho$ values (or higher IPTG concentration values) as the population size decreases. 
In particular, the upper $\rho$ limit of the bistability region as computed from the deterministic CPB model is located at $\rho=0.143$, while the corresponding values for $N=1000$ and 500 cells are $\rho=0.140$ and 0.138, respectively. 
The lower $\rho$ limit of the bistability region also shifts towards lower values for smaller size populations, $\rho=0.085$ for the deterministic CPB model and $\rho=0.084$, 0.082 for $N=1000$ and 500 cells, respectively. 
For noise reduction purposes, we performed 1000 copies of CNMC simulations for the case of $N=1000$ cells and 5000 copies for the case of $N=500$ cells.
The short time interval $\tau$ which reports the discrete evolution of coarse variables is $\tau=0.2$ for all cases.

In Fig.\ref{fig:m_eq_2_f_eq_05}(b), we depict three co-existing coarse steady state solutions (normalised with respect to the average intracellular content, $\left< x \right>$) as obtained from CNMC simulations with $N=1000$ cells for the same parameter value, $\rho=0.096$.
The solid line distributions correspond to the stable upper and lower branch solutions with significantly different average phenotype, i.e., $\left< x \right> =0.62$ for the upper solution branch and $\left< x \right>=0.038$ for the lower solution branch.
The dashed line depicts the unstable steady state solution, which can be tracked only computationally through the application of the arc-length parameter continuation algorithm.
The unstable solution has an average of $\left< x \right>=0.1$. 

\subsection{Effect of partitioning asymmetry during division}
\label{sec:asymmetry}

\begin{figure}[h!]
\begin{center}
  \includegraphics[width=0.65\textwidth]{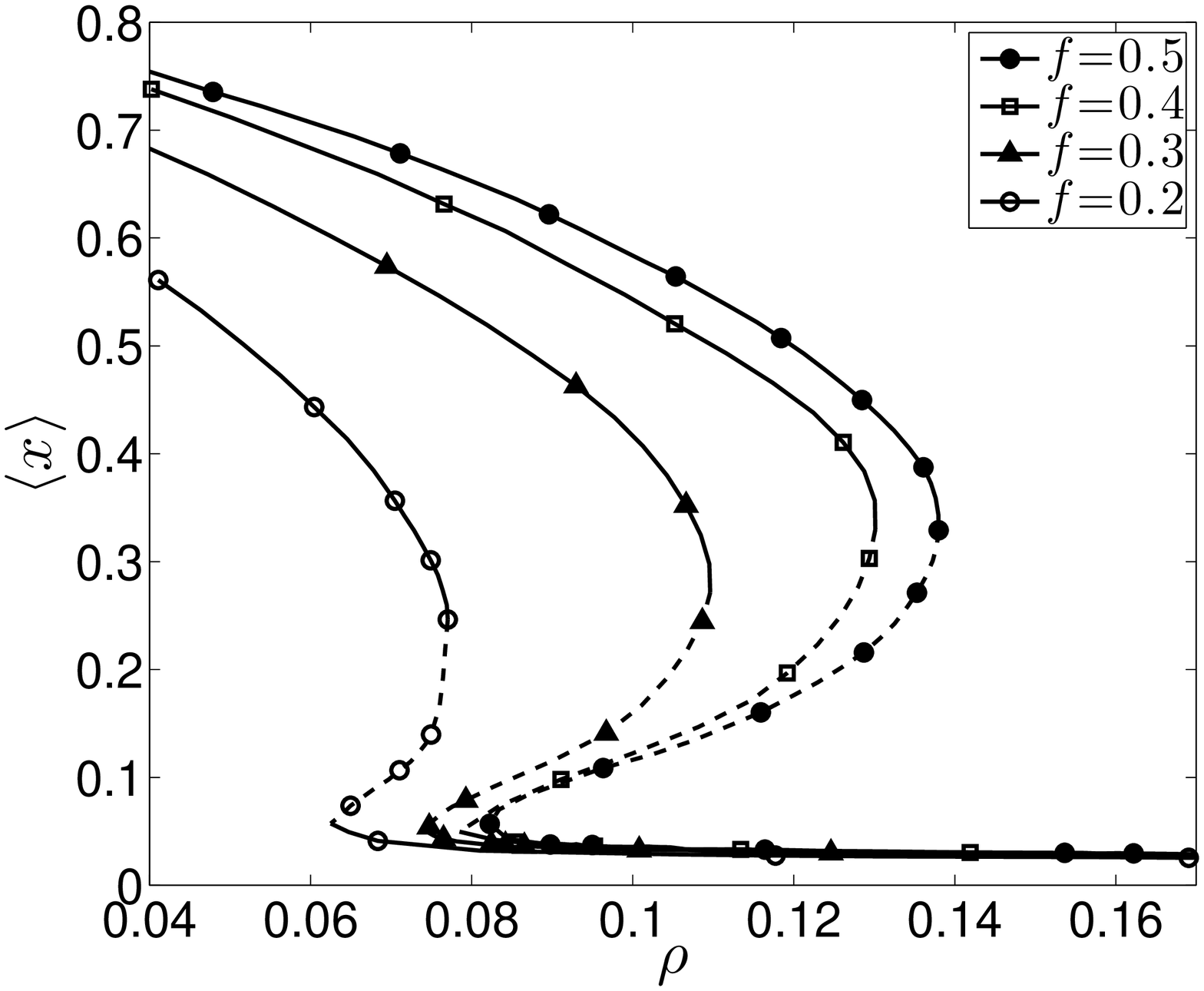}
  \end{center}
\caption{Effect of the partitioning asymmetry factor, $f$, in the average
expression of the {\it lac}Y gene steady state ($\left< x \right>$), as a function of
the inverse IPTG concentration $\rho$.
The line with full circles corresponds to $f=0.5$;
the line with open squares corresponds to $f=0.4$;
the line with full triangles corresponds to $f=0.3$ and the line with open circles to $f=0.2$.
Solid and dashed lines correspond to stable and unstable steady state solutions, respectively. 
The following set of parameter values is used: $m=2$, $\pi=0.03$, and $\delta=0.05$.}
\label{fig:f_comparison}       % Give a unique label
\end{figure}

The effect of partitioning asymmetry is illustrated in Fig.~\ref{fig:f_comparison} for a population of $N=500$ cells.
One can clearly observe the shrinking of the bistability interval for lower $f$ values, suggesting that an increase of
asymmetry in partitioning leads to the extinction of the bistability region.
In particular,  while the bistability region for symmetric partitioning ($f=0.5$) lies within the interval $\rho \in [0.082, 0.138]$, it shrinks to the interval of $\rho \in [0.063, 0.077]$ for $f=0.2$.
A similar behavior is also observed for larger population sizes, as well as for the deterministic CPB model.

\begin{figure}[h!]
\begin{center}
  \includegraphics[width=0.65\textwidth]{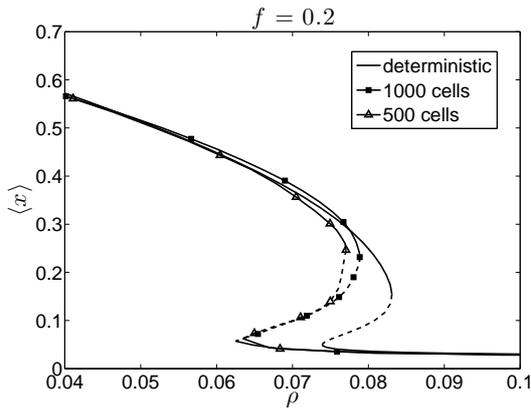}
  \end{center}
\caption{Effect of the population size on the average
expression of the {\it lac}Y gene steady state ($\left< x\right>$), as a function of
the inverse IPTG concentration $\rho$ when the asymmetry partitioning value is $f=0.2$.
The black line corresponds to the deterministic cell population balance model, the line with full squares corresponds to CNMC simulations with $N=1000$ cells,
and the line with open triangles corresponds to CNMC simulations with $N=500$ cells.
Solid and dashed lines correspond to stable and unstable steady state solution branches. 
Parameter set values: $m=2$, $\pi=0.03$, and $\delta=0.05$.}
\label{fig:m_eq_2_f_eq_02}       % Give a unique label
\end{figure}

Interestingly enough, the effect of population size has increasing significance for more asymmetric partitioning mechanisms.
Indicatively, we refer the reader to Fig.~\ref{fig:m_eq_2_f_eq_02}, in which the asymmetry partitioning value is $f=0.2$.
The bistability region for the deterministic case extends over the interval $\rho \in [0.074, 0.083]$, whereas the bistability region when a population of $N=1000$ cells is simulated extends over the
interval $\rho \in [0.063, 0.079]$.

\begin{figure}[h!]
\begin{center}
\begin{tabular}{cc}
  \includegraphics[width=0.5\textwidth]{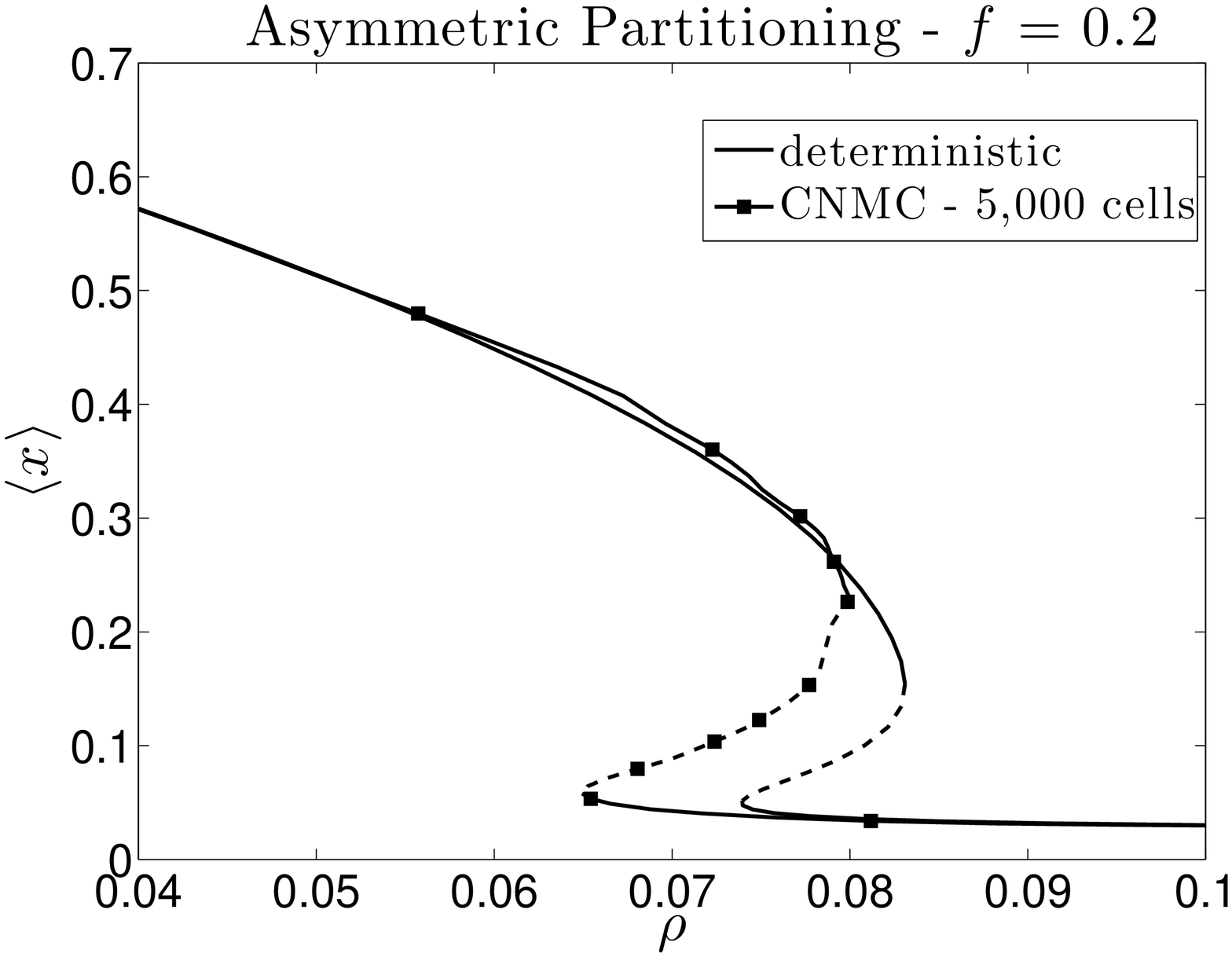} &
  \includegraphics[width=0.5\textwidth]{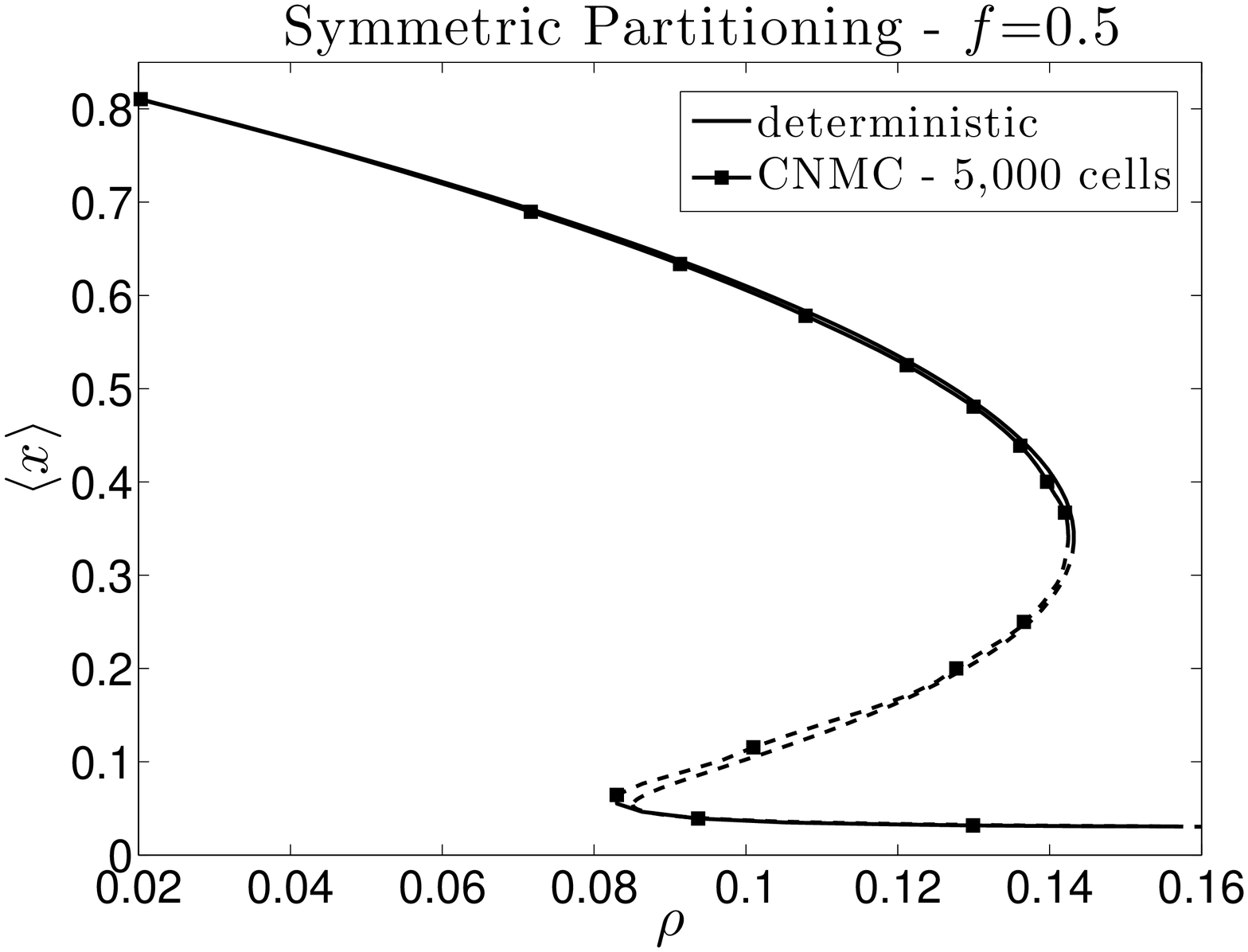} \\
  (a) & (b) 
  \end{tabular}
  \end{center}
\caption{Dependence of the average {\it lac}Y content on the inverse concentration of IPTG ($\rho$) for (a) asymmetric partitioning, $f=0.2$, and (b) symmetric partitioning, $f=0.5$. 
The black line corresponds to the deterministic CPB model and the line with full squares to stochastic CNMC simulations with 5,000 cells (average of 500 copies).
Solid and dashed lines correspond to stable and unstable steady state solutions.
Parameter set values: $\pi=0.03$, and $\delta=0.05$.}
\label{fig:asymmetry_effect}       % Give a unique label
\end{figure}

Large discrepancies between the stochastic and deterministic approaches are also observed in larger size cell populations when the partitioning mechanism is asymmetric.
In Fig.~\ref{fig:asymmetry_effect}(a), we present the dependence of the average {\it lac}Y content on the parameter $\rho$ as obtained from stochastic CNMC simulations with 5,000 cells and the direct CPB model for $f=0.2$.
The stochastic simulations predict a significant shift of the bistability region towards smaller $\rho$ values.
However, when the partitioning mechanism is considered to be symmetric ($f=0.5$) these discrepancies practically vanish (see Fig. ~\ref{fig:asymmetry_effect}(b)).

\subsection{Effect of cell division sharpness}
\begin{figure}[h!]
\begin{center}
  \includegraphics[width=0.65\textwidth]{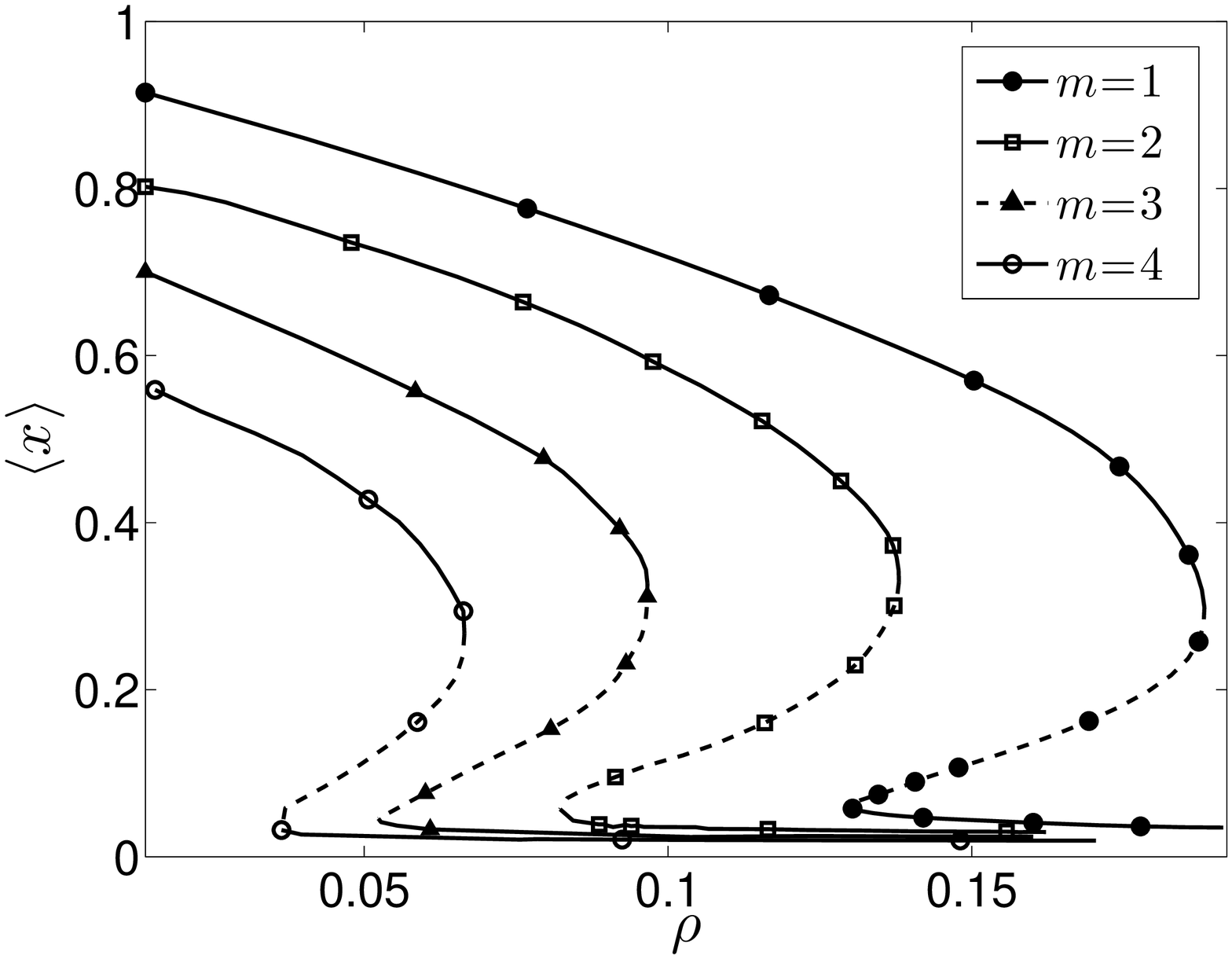}
  \end{center}
\caption{Effect of the division rate sharpness (quantified by $m$) in the average
expression of the {\it lac}Y gene steady state ($\left<x\right>$), as a function of
the inverse IPTG concentration $\rho$.
Higher $m$ values correspond to sharper division rates.
The line with full circles corresponds to $m=1$;
the line with open squares corresponds to $m=2$;
the line with full triangles corresponds to $m=3$ and the line with open circles to $m=4$.
Solid and dashed lines correspond to stable and unstable steady state solutions, respectively.
The following set of parameter values is used: $f=0.5$, $\pi=0.03$, and $\delta=0.05$.}
\label{fig:m_comparison}       % Give a unique label
\end{figure}

\begin{figure}[h!]
\begin{center}
  \includegraphics[width=0.65\textwidth]{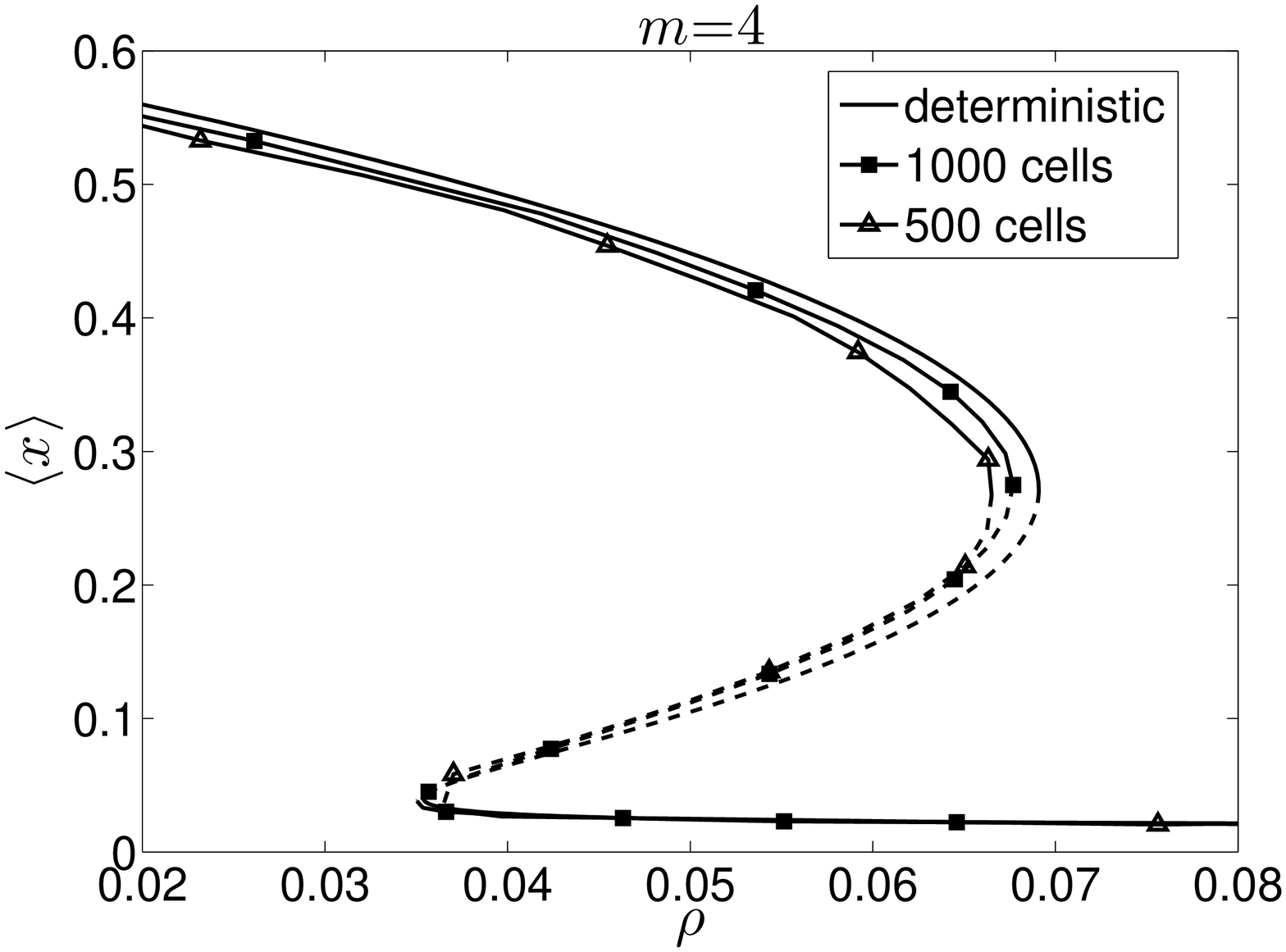}
  \end{center}
\caption{Effect of the population size on the average
expression of the {\it lac}Y gene steady state ($\left< x \right>$), as a function of
the inverse IPTG concentration $\rho$ when cell division rate is sharp, $m=4$.
The black line corresponds to the deterministic cell population balance model, the line with full squares corresponds to CNMC simulations with $N=1000$ cells,
and the line with open triangles corresponds to CNMC with $N=500$ cells.
Solid and dashed lines correspond to stable and unstable steady state solutions, respectively.
Parameter set values: $f=0.5$, $\pi=0.03$, and $\delta=0.05$.}
\label{fig:m_eq_4_f_eq_05}       % Give a unique label
\end{figure}

The effect of division rate sharpness quantified by the parameter $m$ is presented in Fig.~\ref{fig:m_comparison}, when simulations of $N=500$ are executed with the CNMC algorithm.
Larger $m$ values correspond to higher cell division rates. 
The bistability region clearly shrinks and shifts towards higher IPTG concentration values for increasing $m$.
The same behaviour has been also observed when considering the deterministic CPB model \cite{Kavousanakis:2009}.
Here, the effect of cell division sharpness when we compare the deterministic model with the stochastic simulations is not as prominent as 
in the case of the effect of partitioning asymmetry parameter, $f$. 
The larger discrepancies in the phenotypic population behaviour as predicted from the deterministic and CNMC modelling is observed for sharp division rates.
Indicatively, we present the bifurcation diagrams obtained from the deterministic CPB and the stochastic CNMC model with $N=500$ cells for $m=4$ in Fig.~\ref{fig:m_eq_4_f_eq_05}.
The bistability region for the deterministic model is located within the interval $\rho \in [0.035,0.069]$, while for the CNMC model lies within the interval $\rho \in [0.036,0.067]$.

\subsection{Noise driven phenotypic switching}

\begin{figure}[h!]
\begin{center}
  \includegraphics[width=0.65\textwidth]{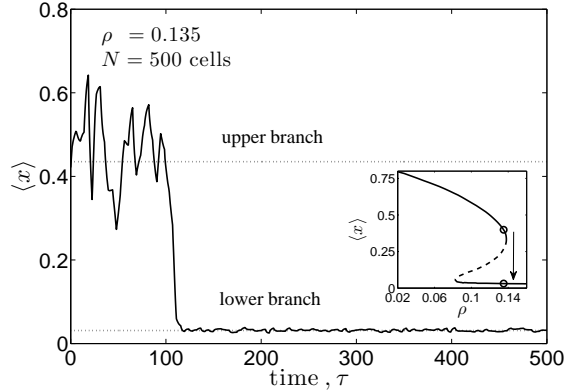}
  \end{center}
\caption{One stochastic simulation starting from a coarse steady state solution at $\rho=0.135$ with $\left< x \right> =0.4$ (upper branch).
Due to stochastic noise the cell population jumps towards the lower stable branch ($\left< x \right>=0.03$) after a long dimensionless time elapse   ($\tau \approx 110$).
The inset depicts the bifurcation diagram of coarse steady solutions as a function of the parameter $\rho$, and with open circles the corresponding stable steady states at $\rho=0.135$.
Parameter set values: $f=0.5$, $\pi=0.03$, $\delta=0.05$, $N=500$ cells, $\rho=0.135$.}
\label{fig:jumps}       % Give a unique label
\end{figure}

\begin{figure}[h!]
\begin{center}
  \includegraphics[width=0.65\textwidth]{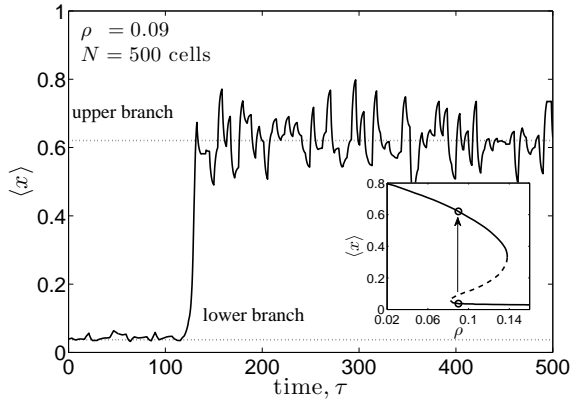}
  \end{center}
\caption{One stochastic simulation starting from a coarse steady state solution at $\rho=0.09$ with $\left< x \right> =0.037$ (lower branch).
Due to stochastic noise the cell population jumps towards the upper stable branch ($\left< x \right>=0.62$) after a long dimensionless time elapse   ($\tau \approx 130$).
The inset depicts the bifurcation diagram of coarse steady solutions as a function of the parameter $\rho$, and with open circles the corresponding stable steady states at $\rho=0.09$.
Parameter set values: $f=0.5$, $\pi=0.03$, $\delta=0.05$, $N=500$ cells, $\rho=0.09$.}
\label{fig:jumps_inverse}       % Give a unique label
\end{figure}

The determination of bistability region provides useful insight for the understanding of some non-trivial behaviour, which can be observed in cell populations. 
In particular, stochastic noise can cause rapid phenotypic transitions when the IPTG concentration lies at the proximity of critical turning point values. 
In Fig.~\ref{fig:jumps} we present an illustrative example of a (single copy) simulation, where the initial condition is a ``coarse'' steady state solution computed for $\rho=0.135$.
The average phenotype of cells  fluctuates around the value of $\left< x \right> \approx 0.4$ for a long time period ($\tau \approx 110$) and then switches towards a low expression level of {\it lac}Y ($\left< x \right>=0.03$).
A reverse phenotypic switch from low to large $\left< x \right>$ values is presented in Fig.~\ref{fig:jumps_inverse}.
The extracellular inducer $\rho=0.09$ value is at the proximity of the left turning point.
After the time elapse of a large period, during which the cell population exhibits low level {\it lac}Y phenotype, stochastic noise drives the system to states of high average expression levels. 
The time at which the phenotypic switch occurs is random.

\section{Discussion}
\label{sec:conclusions}

We presented a multiscale methodology for the computation of the macroscopic steady state behaviour of cell populations, which carry the 
{\it lac} operon genetic network, and are simulated with stochastic fine-scale models. 
In particular, we employ the CNMC model presented in Mantzaris (2006), which incorporates the effect of random events during division and describes with sufficient accuracy the behaviour of small size cell populations.
Despite its accuracy, the stochastic modelling approach lacks efficiency when a coarse-level analysis is required, due to significant computational requirements.
In addition, for systems which exhibit solution multiplicity within a range of parameter values, the strategy of performing solely direct temporal simulations is not the appropriate one, since a very large number of different initial states need to be tested in order to drive the system to different basins of attraction.
A more systematic analysis is applicable when alternative multiscale techniques are applied.

In this work, we adopt techniques from the ``equation-free'' framework \cite{Kevrekidis:2003,Kevrekidis:2004}, which enables the performance of numerical tasks at a systems-level, such as the computation of steady state solutions and the performance of bifurcation analysis at a coarse level of description.
We compute both stable and unstable macroscopic steady state solutions, and determine with accuracy the critical extracellular inducer concentration values (limits of bistability), which signal transitions between the different {\it lac}Y expression levels.
Bifurcation diagrams are computed for different sets of parameter values, in order to examine the effect of partitioning asymmetry of the intracellular content during division, as well as the division rate sharpness on the range of bistability. 
When the partitioning mechanism is highly asymmetric, a shift towards higher IPTG values, as well as shrinking of the bistability range is observed.
This effect was also demonstrated in \cite{Kavousanakis:2009} for the deterministic CPB model case. 
A comparison between the stochastic model and the deterministic CPB model shows that stochasticity tends to shift the bistability range towards higher IPTG concentration values.
The effect of random events during division becomes increasing when CNMC simulations are performed with low size cell populations, and we demonstrated that in the case of highly asymmetric partitioning the discrepancies between stochastic and deterministic models are significant.
Furthermore, stochastic simulations can predict phenotypic switching between different {\it lac}Y expression levels, and we illustrated such transitions at the neighbourhood of the bistability upper and lower limits. 
Such non-trivial dynamical behaviour cannot be predicted by deterministic models
In this work, we focused on the effect of heterogeneity, originating from the uneven distribution of intracellular components of mother cells to their offsprings and random events during division. 
A second source of heterogeneity stems from the fact that intracellular chemical reactions are regulated by small concentrations of chemical species (regulatory molecules) \cite{Alberts:1994:MBC}.
Experimental studies \cite{Blake:2003:NEG,Elowitz:2002:SGE} have demonstrated that the process of gene expression on small synthetic genetic networks is stochastic.
A modelling approach that incorporates the effect of single-cell intrinsic noise effects is presented in Mantzaris (2007), where the deterministic reaction rate $R(x)$ (see Eq.~(\ref{eq:react}),) which describes the evolution of the intracellular species, is replaced by a Langevin equation \cite{Miller:1978:TO}.
The region of bistability has been determined for the single-cell stochastic model, however a systematic analysis of the intrinsic noise effect on the population-level phenotype has not yet been performed;
such an analysis can be performed by applying the equation-free methodology. 
Finally, we report that the presented coarse-grained analysis can be trivially adjusted for cell populations, which carry different genetic network architecture, such as the genetic toggle switch \cite{Gardner:2000:GTC}, which also exhibits bistable behaviour, and switching between the co-existing stable states is possible through thermal or chemical induction.

%\begin{acknowledgements}
%\end{acknowledgements}

\bibliographystyle{spbasic}      % basic style, author-year citations


\end{document}